\documentclass[a4paper,12pt]{article}
\def\al{\alpha}
\def\be{\beta}
\def\ga{\gamma} \def\Ga{\Gamma}
\def\ep{\epsilon}
\def\lam{\lambda}

 \def\calH{{\cal H}} 
 \def\calK{{\cal K}} \def\calL{{\cal L}}
  \def\calO{{\cal O}}
  
 \def\calT{{\cal T}}

\def\del        {  \partial  }
\def\half       {  {1\over 2}  }

\def\ie         {  {\it i.e.}      }

\def\comma          {\, ,}
\def\period         {\, .}
\def\lsim    {\lower .65ex \hbox{\ $\stackrel{<}{\sim}$\ } }
\def\gsim    {\lower .65ex \hbox{\ $\stackrel{>}{\sim}$\ } }
\def\com#1#2   { \left[#1, #2\right]} 
\def\acom#1#2  {\left\{ #1,#2\right\}}
\def\bra#1     {\langle #1 |}
\def\ket#1     {| #1 \rangle}
\def\slash#1{{\ooalign{\hfil/\hfil\crcr$#1$}}} 
\def\vecii#1#2      {  \left(\begin{array}{c}#1\\#2\end{array}\right)  }
\def\veciii#1#2#3   {  \left(\begin{array}{c}#1\\#2\\#3\end{array}
                     \right)  }
\def\veciv#1#2#3#4  {  \left(\begin{array}{c}#1\\#2\\#3\\#4
                                 \end{array}\right)  }
\def\vecfv#1#2#3#4#5 {  \left(\begin{array}{c}#1\\#2\\#3\\#4\\#5
                                 \end{array}\right)  }

\def\matrixii#1#2#3#4            {  \left(\begin{array}{cc}#1&#2\\#3&#4
                                       \end{array}\right) }
\def\matrixiii#1#2#3#4#5#6#7#8#9 {  \left(\begin{array}{ccc}#1&#2&#3\\
                                     #4&#5&#6\\#7&#8&#9\end{array}
                               \right)  }
\def\mativ#1#2#3#4               {  \left(\begin{array}{cccc}
                                       #1\\#2\\#3\\#4\end{array}\right) }
\def\matv#1#2#3#4#5              {  \left(\begin{array}{ccccc}
                                     #1\\#2\\#3\\#4\\#5\end{array}
                              \right)  }
\def\eqabegin         {  \begin{eqnarray}  }
\def\eqaend           {  \end{eqnarray}  }
\def\nn               {  \nonumber  }
\def\bracetwo#1#2     {  \left\{ \begin{array}{l} #1 \\ #2 \end{array}
                         \right.  }
\def\bracetwocases#1#2#3#4  {   \left\{ \begin{array}{ll} #1 &
                                 \qquad #2 \\
                                 #3 & \qquad #4 \end{array} \right.  }
\def\bracebegin#1     {  \left\{ \begin{array}{#1}   }
\def\braceend         {  \end{array}\right.   }
\def\parn              {  \par\noindent }

\def\parmedskip        {  \par\medskip  }

\def\parag#1           {\paragraph{#1} \mbox{ }\parmedskip\noindent}
\def\msection#1      {  \begin{center} \section{#1} \end{center}   }
\def\nsection#1      {  \let\boldface\bf \def\bf{} \section{#1}
                           \let\bf\boldface   }
\def\mnsection#1     {  \begin{center} \nsection{#1} \end{center}  }
\def\capsection#1    {  \let\boldface\bf \def\bf{\sc} \section{#1}
                           \let\bf\boldface   }
\def\mcapsection#1   {  \begin{center} \capsection{#1} \end{center} }

\newcommand{\nullify}[1]{}

\def\papertitlepage{\baselineskip 3.5ex \thispagestyle{empty}}
\def\Title#1{\baselineskip 1cm \vspace{1.5cm}\begin{center}
 {\Large\bf #1} \end{center} 
\vspace{0.5cm}}
\def\Authors#1{\begin{center} {\it #1} \end{center}}
\def\Abstract{\vspace{1.0cm}\begin{center} {\large\bf Abstract} 
           \end{center} \par\bigskip}
\def\Komabanumber#1#2#3{\hfill \begin{minipage}{4.2cm} UT-Komaba #1
              \parn #2 
              \parn #3 \end{minipage}}
\renewcommand{\thefootnote}{\fnsymbol{footnote}}
\renewenvironment{thebibliography}{\pagebreak[3]\par\vspace{0.6em}
\begin{flushleft}{\large \bf References}\end{flushleft}
\vspace{-1.0em}

\begin{enumerate}\if@twocolumn\baselineskip=0.6em\itemsep -0.2em
\else\itemsep -0.2em\fi\labelsep 0.1em}{\end{enumerate}}

\def\xiinv{\xi^{-1}}
\def\Vinv{V^{-1}}
\def\Ginv{G^{-1}}
\def\Uhat{\hat{U}}
\def\Tzero{T^{(0)}}
\def\Tone{T^{(1)}}

\def\atil{\tilde{a}}
\def\btil{\tilde{b}}
\def\acth{\Acute{\theta}}
\def\mathd{d}
\def\mathi{i}

\def\Poisson[#1,#2]{\{#1,\,#2\}_{P}}
\def\Poissonss[#1,#2]{\{#1(\sigma),\,#2(\sigma')\}_{P}}
\def\Dirac[#1,#2]{\{#1,\,#2\}_{D}}
\def\Diracss[#1,#2]{\{#1(\sigma),\,#2(\sigma')\}_{D}}
\def\DDirac[#1,#2]{\{#1,\,#2\}_{D}}
\def\DDiracss[#1,#2]{\{#1(\sigma),\,#2(\sigma')\}_{D}}

\usepackage{graphicx}
\usepackage{latexsym}
\usepackage{amsmath}
\usepackage{amssymb}
\usepackage{wick}

\renewenvironment{thebibliography}{\pagebreak[3]\par\vspace{0.6em}
\begin{flushleft}{\large \bf References}\end{flushleft}
\vspace{-1.0em}

\begin{enumerate}\if@twocolumn\baselineskip=0.6em\itemsep -0.2em
\else\itemsep -0.2em\fi\labelsep 0.1em}{\end{enumerate} }
\def\adot{{\dot{a}}}
\def\bdot{{\dot{b}}}

\def\xdot{\dot{x}}

\def\aldot{{\dot{\al}}}
\def\bedot{{\dot{\be}}}
\def\gadot{{\dot{\ga}}}
\def\deltadot{{\dot{\delta}}}

\def\thtildot{\dot{\tilde{\theta}}}
\def\Qtil{\tilde{Q}}
\def\lamtil{\tilde{\lambda}}

\def\Qhat{\hat{Q}}

\def\thbar{\bar{\theta}}

\def\Thbar{\bar{\Theta}}
\def\wtil{\tilde{w}}
\def\lamtil{\tilde{\lam}}
\def\Qtil{\tilde{Q}}

\def\thtil{\tilde{\theta}}

\def\ptil{\tilde{p}}
\def\ktil{\tilde{k}}
\def\dtil{\tilde{d}}
\def\Ktil{\tilde{K}}
\def\ptil{{\tilde{p}}}

\def\Dtil{\tilde{D}}
\def\Phitil{\tilde{\Phi}}
\def\th{\theta}
\def\Th{\Theta}
\def\sig{\sigma}
\def\sigp{\sigma'}

\def\pslash{\slash{p}}

\def\pplus{{p^+}}

\def\deltassp{\delta(\sigma-\sigma')}

\def\Dcom#1#2{\bigl\{#1,\,#2\bigr\}_D}

\def\Pcom#1#2{\bigl\{#1,\,#2\bigr\}_P}

\def\supzero{{(0)}}
\def\supone{{(1)}}

\def\suptwo{{(2)}}
\def\supthree{{(3)}}
\def\ginv{g^{-1}}
\def\vinv{v^{-1}}

\def\xiinv{\xi^{-1}}

\def\mbar{{\bar{m}}}
\def\pbar{{\bar{p}}}
\def\qbar{{\bar{q}}}
\def\agnum{\overline{{\rm gh}}\#   }
\def\ctil{\tilde{c}}
\def\omtil{\tilde{\omega}}
\def\gnum{{\rm gh}\#   }
\def\npb#1{Nucl. Phys. {\bf B#1}}

\def\jhep#1{JHEP {\bf #1}}
\def\hepth#1{ hep-th/#1}
\def\plb#1{Phys. Lett. {\bf B#1}}

\begin{document}
\papertitlepage
\vspace*{0cm}
\Komabanumber{06-1}{hep-th/0603004}{March, 2006}
\Title{Towards Pure Spinor Type Covariant Description of \\
Supermembrane \\
{\large --- An Approach from the Double Spinor Formalism ---}} 
\Authors{{\sc Yuri Aisaka\footnote[2]{yuri@hep1.c.u-tokyo.ac.jp} 
 and Yoichi Kazama\footnote[3]{kazama@hep1.c.u-tokyo.ac.jp}
\\ }
\vskip 3ex
 Institute of Physics, University of Tokyo, \\
 Komaba, Meguro-ku, Tokyo 153-8902 Japan \\
  }
\baselineskip .7cm

\numberwithin{equation}{section}
\numberwithin{figure}{section}
\numberwithin{table}{section}

\parskip=0.9ex

\Abstract

In a previous work, we have constructed a reparametrization invariant 
worldsheet action from which one can derive the super-Poincar\'e covariant 
pure spinor formalism for the superstring at the fully quantum level.
The main idea was the doubling of the spinor degrees of freedom in the
Green-Schwarz formulation together with the introduction of a new 
compensating local fermionic symmetry. In this paper, we extend 
this ``double spinor" formalism to the case of the supermembrane 
in 11 dimensions at the classical level. 
The basic scheme works in parallel with the string case and we 
are able to construct the closed algebra of first class constraints
which governs the entire dynamics of the system. 
A notable difference from the string case is that this algebra 
is first order reducible and the associated BRST operator must be 
constructed accordingly. 
The remaining problems which need to be solved for the 
quantization will also be discussed.

\newpage
\baselineskip 3.5ex
\section{Introduction}  
\renewcommand{\thefootnote}{\arabic{footnote}}
Six years ago, N.~Berkovits opened up a novel perspective 
 for the quantization of the superstring with
 manifest super-Poincar\'e covariance by proposing the so-called 
 pure spinor (PS) formalism~\cite{Berk0001}.
 The basic ingredient of this formalism is the BRST-like 
 operator $Q =\int dz \lam^\al d_\al$, 
 where $d_\al = p_\al + i\del x^m (\ga_m \th)_\al
+\half (\ga^m \th)_\al (\th \ga^m\del \th)$ coincides with the 
 familiar constraint that
 arises  in the conventional Green-Schwartz (GS) formalism and $\lam^\al$ is 
a bosonic chiral spinor playing the role of the associated ``ghost". 
For $Q$ to be regarded as a BRST operator, however, $\lam^\al$ must 
 satisfy a subsidiary constraint. 
With the assumption that all the fields are free, 
one obtains
 the operator product $d_\al(z)d_\be(w) = 2i \ga^m_{\al\be} (\del x_m
-i\th \ga_m \del \th)/(z-w)$ and hence $Q$ becomes nilpotent if and only if 
 $\lam^\al \ga^m_{\al\be} \lam^\be =0$. This, in 10 dimensions, is 
precisely the condition for $\lam^\al$ to be a 
{\it pure spinor} in the sense of Cartan~\cite{Cartan}.

 The striking 
property of this  operator $Q$ is that, despite its simplicity,
  its cohomology correctly reproduces the 
 spectrum of the superstring~\cite{Berk0006}. 
Moreover, together with the field $\omega_\al$
 conjugate to $\lam^\al$, the fields in the theory form 
 a conformal field  theory (CFT) with vanishing central charge, which 
 allows one to make use of the powerful machinery of CFT. $Q$-invariant
 vertex operators were constructed and by postulating an appropriate 
 functional measure the known tree level amplitudes were reproduced 
 in a manifestly covariant manner~\cite{Berk0001,Berk0004}. 
Subsequently, this success was 
 extended to the multi-loop level~\cite{BerkMulti}.
Some explicit supercovariant 
 calculations have been performed at 1 and 2 loops~\cite{BerkTwo}, which agreed with
the  results obtained in the Ramond-Neveu-Schwarz (RNS) formalism~\cite{FMS86,DhokerPhong}.
 Furthermore certain vanishing theorems were proved
 to all orders for the first time, demonstrating the 
 power of this formalism~\cite{BerkMulti}. 
Another advantage of the PS formalism is 
 that it can be coupled to backgrounds including 
 Ramond-Ramond fields in a
 quantizable and covariant way~\cite{Berk0001,PSCurved}, 
 in distinction to the conventional 
 RNS and GS formalisms, where
 one encounters difficulties.

  More recently it has been shown that, 
 with some additional fields, the original PS formalism can be promoted to 
 a new ``topological" formulation~\cite{BerkTopol}\footnote{%
See also~\cite{Nekrasov:2005wg}.
}, where the structure of the loop amplitude
 becomes very similar to the bosonic string, just like in the case of 
 the topological string~\cite{topstrings}.  This structure may shed more light on 
 the deeper understanding of the PS formalism. 
For many other 
 developments, the reader is referred to 
 \cite{BerkOthers}\nocite{Vallilo:2003nx,Berkmembrane,Stonybrook,SE,Trivedi,AK,AK4,Chesterman,Bstates,Oota,Guttenberg:2004ht,Grassi2}--\cite{PSLower}
 and a review \cite{BerkRev}. 

Behind  these remarkable advances, there remained a number of 
 important mysteries 
concerning this formalism: What is the underlying reparametrization 
 invariant worldsheet action and 
what are its symmetries?
How does $Q$ arise as a BRST operator? 
Why are all the fields free? 
How does $\lam^\al$ get constrained and how does one  quantize it? 
Why is the Virasoro constraint absent in $Q$? 
How does one derive the functional measure? 
In summary, the basic problem was to understand 
 the origin of the PS formalism. 

In a previous work~\cite{AK4}, 
 we have given answers to many of the above questions
 by constructing a fundamental reparametrization invariant action 
 from which one can derive the PS formalism 
 at the fully quantum level. As we shall review in Sec.~\ref{sec:DSrev}, 
the basic idea was to add a new spinor degree of freedom $\th^\al$ to the 
 Green-Schwarz action consisting of $x^m$ and $\thtil^\al$, 
in such a way that a compensating local fermionic 
 symmetry appears on top of the usual $\kappa$-symmetry~\cite{Siegel83}. Due to this 
extra symmetry, the physical degrees of freedom remain  unchanged. 
Just as in the usual GS formalism, the standard Hamiltonian analysis
 \`a la Dirac shows that both the first and the second class 
constraints arise, which cannot be separated without breaking  manifest
Lorentz  covariance. Now the advantage of the ``double spinor"
 formalism is that 
this breakdown can be confined to the $\thtil$ sector while the covariance 
 for $x^m$ and $\th^{\alpha}$ remain intact. Then after fixing the $\kappa$-symmetry
for the $\thtil$ sector by adopting the semi-light-cone (SLC) gauge, 
one obtains a closed set of first class constraints which govern 
 the entire dynamics of the theory. This algebra, which is absent in 
 the conventional GS formalism, has its origin in the aforementioned extra 
 local fermionic symmetry and is the most important feature of the
 double spinor formalism.

 Furthermore, by appropriate redefinitions of the 
 momenta, one can construct a basis of fields in which the Dirac brackets 
among them take the canonical ``free field" form. This at the same time 
 simplifies the form of the constraints. 
 The quantization can then be performed by replacing the Dirac bracket
 by the quantum bracket together with  slight quantum modifications of the form of the constraints due to multiple contractions and normal-ordering. 
The quantum first class algebra so obtained precisely matches 
 the one proposed in \cite{BM0412} and  justifies the free-field
 postulate of Berkovits\footnote{%
 This first class algebra can also be obtained by the so-called
 ``BFT embedding method''\cite{GanonaGarcia}}. 

 The nilpotent BRST operator $\Qhat$
 associated with it can easily be constructed in the standard way, with the introduction 
 of {\it unconstrained } bosonic spinor ghosts $\lamtil^\al$ and the 
reparametrization ghosts $(b,c)$ associated with the Virasoro constraint. 
At this stage, $\Qhat$ still contains non-covariant pieces representing the 
part of the degrees of freedom of the gauge-fixed $\thtil^\al$. 
The remarkable fact is that all the unwanted components in $\Qhat$ 
can be removed or cohomologically decoupled  
through a quantum similarity transformation: The Virasoro generator 
 disappears together with the $b,c$ ghosts and the non-covariant remnants 
 of $\thtil^\al$ cancel against a part of the unconstrained $\lamtil^\al$ 
 in such a way that it precisely becomes a pure spinor $\lam^\al$ 
 satisfying the quadratic PS constraint. In this way one finally arrives at
 the Berkovits' expression $Q=\int dz \lam^\al d_\al$. In \cite{AK4} 
it was also shown that the same method can be used to derive 
 the PS formalism for a superparticle 
 in  11 dimensions. 

Now, an obvious  and challenging question arises: Is the above idea 
applicable to the supermembrane in 11 dimensions as well? 

Some years ago, 
the possibility of a  pure spinor type formalism for the supermembrane
 was investigated by Berkovits~\cite{Berkmembrane}.
 Largely based on the requirement that the 
 theory should reduce in appropriate limits to that of 
a 11 dimensional superparticle  and a 10 dimensional type IIA superstring, 
he generalized the conventional
 supermembrane action
 first written down by 
Bergshoeff, Sezgin and Townsend (BST)~\cite{Bergshoeff:1987cm} to include 
 a bosonic spinor $\lam^\al$ and its conjugate $\omega_\al$. This action 
 is invariant under a postulated BRST transformation generated by 
$Q=\int d^2\sig\lam^\al d_\al$,  which is nilpotent if a set of constraints on $\lam^\al$
 are satisfied. In addition to the familiar one 
 $\Bar{\lam} \Ga^M \lam =0$, 
 this set includes further new constraints involving worldvolume derivatives. 
Unlike the case of the superparticle and the superstring, the action 
 is non-linear and the problem of quantization was left unsolved. 
Nonetheless, this pioneering study gave some hope 
 that a covariant quantization of a supermembrane may be possible 
 along the lines of the pure spinor formalism. Our work  to be presented 
 in this paper is another attempt for this challenging task 
 from a different more systematic point of view. 

We will now outline the results of our investigation, which at the same time
 indicates the organization of the paper. 

We begin in Sec.~\ref{sec:DSrev} with
 a  review of how the double spinor formalism works in the case 
 of the superparticle and the superstring. This should help the reader to 
form a clear picture of the basic mechanism, without being hampered by
 the complicated details of the membrane case stemming from the high degree
 of added non-linearity.

The main
analysis for the supermembrane
 case is performed in Sec.~\ref{sec:DSSM}.
The fundamental action we start with in Sec.~\ref{subsec:DSSMaction}
is formally of the same form as the conventional one~\cite{Bergshoeff:1987cm}, 
 except that (i) the spinor variable $\thtil_A$
 is replaced by $\thtil_A-\th_A$ where
 $\th_A$ is the newly introduced spinor and (ii) the membrane coordinate $x^M$
 is replaced by $x^M-i\thbar \Ga^M \thtil$. Due to these modifications, 
 the action acquires an extra local fermionic symmetry, which will play 
 the crucial role. Then, through the usual Dirac analysis, we  obtain 
 in Sec.~\ref{subsec:DSSMconstr} 
 the fundamental constraints of the system. Due to the presence
 of the extra spinor $\th_A$ and its conjugate momentum, there will be an 
 additional fermionic constraint $D_A$ besides the usual one $\Dtil_A$ 
 associated with $\thtil_A$. 
Upon defining the Poisson brackets for the fundamental fields, 
 we compute the algebra of constraints. This reveals, just as in the 
 case of the superstring, a half of $\Dtil_A$ are second class 
and the remaining half are first class. On the other hand, 
the combination $\Delta_A = D_A +\Dtil_A$, which generates 
 the extra fermionic symmetry,  anticommutes with both $D_A$ and 
 $\Dtil_A$. To separate the first and the second class part 
of $\Dtil_A$, we will make use of the light-cone decomposition. Then
 the first class part $\Ktil_\aldot$ can be identified as the generator of 
 the $\kappa$ transformation. Although the computations 
are  much more involved compared to the string case, 
we show that the anticommutator $\acom{\Ktil_\aldot}{\Ktil_\bedot} $
closes into a bosonic 
expression $\calT_{\aldot\bedot}$, which, although somewhat complicated,  
 is  equivalent to the bosonic constraints 
coming from the original worldvolume reparametrization invariance. 
Next in Sec.~\ref{subsec:DSSMSLC}, 
 we perform the semi-light-cone gauge fixing and 
eliminate the $\kappa$-generators as well as the original 
 second class constraints,  by defining the appropriate Dirac brackets. 
We are then left with the remaining fermionic constraints $D_A$ and 
the bosonic constraints $\calT_{\aldot\bedot}$. Direct computation
 of the algebra of these quantities under the Dirac bracket is unwieldy 
 but we found a way to determine it efficiently by indirect means.
 The result is  a conceptually simple first class algebra, 
which governs the entire dynamics of the system. 
A notable difference from the string case, however, is that this algebra
 is first order reducible, namely that there is a linear relation 
 between some of the constraints. The associated BRST operator, therefore, 
 must be constructed according to the general theory~\cite{HenTb}
 applicable to such a situation. This procedure is described in 
Sec.~\ref{subsec:DSSMBRST}. 

Thus, as far as the general scheme of the double spinor formalism is 
 concerned, we have found that it is indeed applicable to the supermembrane
 case as well and produces a BRST operator in which the covariance is 
 retained for the bosonic coordinate $x^M$ and the new spinor $\th$, which 
 is crucial for the would-be pure spinor type covariant formulation. 
Unfortunately, the remaining steps for proper quantization and elimination 
 of the non-covariant remnants 
 present a number of difficulties and at the present stage 
we have not yet been able to obtain complete solutions. 

In preparation 
 for future developments, however,  we will spell out in Sec.~\ref{sec:quant} the nature of these problems and present
 some preliminary investigations. The first problem, 
 discussed in Sec.~\ref{subsec:ffbasis},  is the construction of the basis in which the
 fields become ``free" under the Dirac bracket. In the case of
 the superstring, this problem was solved completely in  closed form,
 which was a crucial ingredient for the justification of
 the free field postulate 
 of Berkovits. For the supermembrane, it gets considerably more complicated. 
Nevertheless, we will show that the desired basis can be explicitly constructed
 in closed form in the case of the usual BST formulation 
without the new spinor $\th$. For the double spinor formalism, 
it is  accomplished as yet partially but the result strongly indicates 
 the existence of such a basis. 
The second problem is that of quantization. Even if such a ``free field" basis 
 is found, the replacement of the Dirac bracket by the quantum bracket 
 is but a part of the quantization procedure. We will discuss 
what should be achieved for a complete quantization and present a 
preliminary analysis. 
 
Finally, in Sec.~5 we will briefly summarize the main points of
 this investigation and discuss future problems. 

Four appendices are provided: Our notations and conventions
 are summarized in Appendix A, useful formulas in the SLC gauge are
 collected in Appendix B, a proof of the equivalence of 
sets of bosonic constraints is given in Appendix C and 
the first order reducibility function is obtained in Appendix D.

\section{Basic idea of the double spinor formalism:\ A review}
\label{sec:DSrev}
Let us 
begin with a review of the double spinor formalism for the case of the 
lower dimensional objects,  which should serve as a reference point
for the more complicated supermembrane case. 
To highlight the essence of the basic idea, we will concentrate on the 
 simpler case of the superparticle and then supplement
 some further technical refinements needed for the superstring. 
\subsection{Superparticle}
To motivate the double spinor formalism, it is useful to 
 first recall the origin of the difficulty of the covariant 
 quantization of a superparticle in the conventional Brink-Schwarz (BS)
 formalism~\cite{BS}. 
In this formulation, a (type I) superparticle in 10 dimensions 
is described by the reparametrization invariant action given by 
\begin{align}
S_{BS} &= \int dt {1\over 2e} \Pi^m \Pi_m  \comma \qquad 
\Pi^m = \xdot^m -i\thtil \ga^m \thtildot \comma \label{BSaction} \nn
\end{align}
where $e$ is the einbein, $\thtil^\al$ is a 16 dimensional Majorana-Weyl 
spinor,  and the Lorentz
 vector index $m$ runs 
 from $0$ to $9$. The generalized momentum 
$\Pi^m$,  and hence the action, is invariant under 
 the global supersymmetry transformation $\delta \thtil^\al = \ep^\al, 
 \delta x^m = i\ep \ga^m \thtil$. In addition, the action is 
 invariant under the $\kappa$-symmetry transformation of the form~\cite{Siegel83}
$\delta \thtil =\Pi_m \ga^n \kappa, \delta x^m = i\thtil \ga^m \delta \thtil, 
 \delta e = 4ie \thtildot \kappa$, where $\kappa_\al$ is 
 a local fermionic parameter. 

From the definitions of the momenta $(p_m, \ptil_\al, p_e)$
 conjugate to 
$(x^m, \thtil^\al, e)$ respectively,
 one obtains the following two primary constraints:
\begin{align}
\dtil_\al 
\equiv \ptil_\al -ip_m(\ga^m \thtil)_\al =0\comma \qquad p_e =0 \period
\end{align}
Then, the consistency under the time development generated by the 
 canonical Hamiltonian $H = (e/2)p^2$ requires the additional 
 bosonic constraint 
\begin{align}
T &\equiv \half p^2 =0 \period
\end{align}
Hereafter, we will drop $(e,p_e)$ by choosing the gauge $e=1$.
Then, 
taking the basic Poisson brackets as 
\begin{align}
\Pcom{x^m}{p_n} &= \delta_n^m \comma \qquad 
\Pcom{\ptil_\al}{\thtil^\be} = -\delta_\al^\be \comma 
\end{align}
the remaining constraints form  the algebra
\begin{align}
\Pcom{\dtil_\al}{\dtil_\be} &= 2i \pslash_{\al\be} \comma \qquad 
\Pcom{\dtil_\al}{T} = \Pcom{T}{T} =0  \period \label{constalgp}
\end{align}
This is where the necessity of non-covariant treatment 
becomes evident: On the constrained surface $p^2=0$, the quantity 
$\pslash$ has rank 8, indicating that eight of the $\dtil_\al$ are 
of second class and the remaining eight are of first class. 
Since there is no eight-dimensional representation of the Lorentz group, 
manifest covariance must be sacrificed in order to separate these two types 
 of constraints. 

To perform the separation, one employs the $SO(8)$ decomposition 
\begin{align}
p^m &= (p^+, p^-, p^i)\comma \quad p^\pm = p^0 \pm p^9 
\comma \quad i=1\sim 8 \comma \\
\dtil_\al &= (\dtil_a, \dtil_\adot) \comma \qquad a, \adot =1\sim 8 \period
\end{align}
Then from (\ref{constalgp}) it is easily checked that 
 $\dtil_a$ are of second class and the combinations
\begin{align}
\Ktil_\adot &\equiv \dtil_\adot -{p_i \over p^+} \ga^i_{\adot b} 
\dtil_b \comma 
\end{align}
which generate the $\kappa$-transformation, form a set of first
 class constraints together with $T$. 
The rest of the procedure is standard: The second class constraints
are handled by introducing the Dirac bracket, while the first class 
 constraints can be treated by adopting the (semi-)light-cone gauge. 
After that the quantization can be performed in a straightforward manner. 

The well-known analysis recalled above clearly shows that, as long as one 
 employs a single spinor $\thtil^\al$, there is no way to {\it derive}
a covariant quantization scheme such as the pure spinor formalism of our 
 interest. As we demonstrated in a previous work, this problem can be 
 overcome by introducing an additional spinor $\th^\al$, together with 
a new compensating local fermionic symmetry to keep the physical content 
 of the theory intact. 

The new action is formally the same as the Brink-Schwarz action 
(\ref{BSaction}), except that $\thtil$ and $x^m$ are replaced as 
\begin{align}
\thtil &\rightarrow \Th \equiv \thtil -\th\comma \qquad 
x^m \rightarrow 
y^m \equiv x^m -i\th \ga^m \thtil 
\period \label{modification}
\end{align}
As we keep the new spinor $\th$  till the end while $\thtil$ will be 
eliminated, the global supersymmetry 
 transformations are taken as $\delta \th = \ep, \delta \thtil =0, 
\delta x^m =i\ep \ga^m \th$. Because $\th$ is introduced in 
 the simple difference $\thtil -\th$, there arises an apparently
 trivial local fermionic
 invariance under $\delta \th =\chi, \delta \thtil =\chi$, where $\chi$ is 
 a local fermionic parameter. If we gauge-fix $\th$ to be zero using this
 symmetry, we get back the original Brink-Schwarz action. 
It is easily checked that the $\kappa$-symmetry for $\thtil$
 (with $\delta \th\equiv 0$) remains intact. 

The standard Dirac analysis generates the following constraints:
\begin{align}
\Dtil_\al &= \ptil_\al -i(\pslash \thtil)_\al  =0 \comma \qquad 
D_\al = p_\al -i(\pslash(\theta-2\thtil))_\al =0  
\comma \\
T &= \half p^2 =0  \period
\end{align}
$D_\al$ is a new constraint associated with $\th^\al$. 
It is more convenient to replace it with the linear combination 
 $\Delta_\al \equiv D_\al + \Dtil_\al$, which can be identified 
 as the generator of the extra local fermionic symmetry. Since
$\Delta_\al$ can be easily checked to Poisson anti-commute with 
$\Dtil_\al$ and with itself, the only non-vanishing bracket is 
\begin{align}
\Pcom{\Dtil_\al}{\Dtil_\be} &= 2i \pslash_{\al\be} \period
\end{align}
The situation being exactly the same as in the BS formalism, we must 
 employ the light-cone decomposition to identify $\Dtil_a$  as the second 
class and the $\kappa$-generator $\Ktil_\adot = \Dtil_\adot 
 -(p^i/\pplus) \ga^i_{\adot b} \Dtil_b$ as the first class part 
 of $\Dtil_\al$. They (anti-)commute with each other and with $T$ and 
satisfy the relations 
\begin{align}
\Pcom{\Dtil_a}{\Dtil_b} &= 2i\pplus \delta_{ab} 
\comma \qquad 
\Pcom{\Ktil_\adot}{\Ktil_\bdot} = -4i {T \over \pplus} \delta_{\adot\bdot}
\period
\end{align}
Now by imposing the SLC gauge $\thtil_\adot=0$, 
$\Ktil_\adot$'s are turned into second class and,  together with $\Dtil_a$, 
are handled by the use of the appropriate Dirac bracket $\Dcom{\ \ }{\ \ }$. Upon this
 step, the remaining part of $\thtil$, namely $\thtil_a$, becomes 
self-conjugate: With a slight rescaling, we have 
$S_a \equiv \sqrt{2\pplus}\, 
\thtil_a$ satisfying  $\Dcom{S_a}{S_b} = i\delta_{ab}$. 

The crucial difference from the BS formalism is that, under the 
 Dirac bracket,  {\it we still have 
a non-trivial  first class algebra formed by $T$ and 
 $D_\al$} (which is the same 
 as $\Delta_\al$ since $\Dtil_\al$ has been set strongly to zero). 
It reads 
\begin{align}
\Dcom{D_\adot}{D_\bdot} &= -4i {T \over \pplus} \delta_{\adot\bdot}
\comma \qquad \mbox{rest} =0 \period \label{classalg}
\end{align}
This is identical in form to the one satisfied by $\Ktil_\adot$
 above (under the Poisson bracket) and shows that, through the 
 new local fermionic symmetry,  the content of  
the $\kappa$-symmetry for $\thtil$ is transferred to the sector involving the 
 new spinor $\th$. 

The quantization is performed by replacing 
 the Dirac bracket by the quantum bracket. With a slight rescaling of 
 fields, we can set $\com{x^m}{p_n} = i\delta_n^m, \acom{p_\al}{\th^\be}
 = \delta_\al^\be, \acom{S_a}{S_b} = \delta_{ab}$. 
The classical algebra (\ref{classalg}) then turns into the quantum algebra
\begin{align}
\acom{D_\adot}{D_\bdot} &= -4i {T \over \pplus} \delta_{\adot\bdot}
\comma \qquad \mbox{rest} =0 \comma \label{quantalg}
\end{align}
with 
\begin{align}
D_a &= d_a + i\sqrt{2\pplus}\, S_a \comma \quad 
D_\adot = d_\adot +i \sqrt{{2\over \pplus}}\, p^i \ga^i_{\adot b}S_b 
\comma 
\end{align}
where we have separated for convenience the spinor covariant 
 derivative $d_\al \equiv p_\al + (\pslash \th)_\al$ for the $\th$ sector. 

It is now straightforward to construct the nilpotent 
BRST operator associated with  the above first class algebra. It reads
\begin{align}
\Qhat &= \lamtil^\al D_\al + {2\over \pplus} \lamtil_\adot \lamtil_\adot b
 + cT \comma 
\end{align}
where $\lamtil^\al =(\lamtil_a, \lamtil_\adot)$ is an {\it unconstrained} 
bosonic spinor ghost and $(b,c)$ are the usual fermionic ghosts
 satisfying $\acom{b}{c} =1$. Note that the familiar important relation 
 $\acom{\Qhat}{b} =T$ holds. 

The remaining task is to show that this $\Qhat$
 has the same cohomology as the Berkovits' $Q=\lam^\al d_\al$, with 
the constraint 
 $\lam \ga^m \lam =0$. This can be done by  suitable quantum similarity 
 transformations, which preserve the nilpotency and the cohomology.
 First, to remove $T, c$ and $b$, we introduce an auxiliary field $l_\adot$
 with the properties $\lamtil_\adot l_\adot =1, l_\adot l_\adot =0$ and 
form the following operator $b_B$
\begin{align}
b_B &\equiv -{\pplus \over 4} l_\adot D_\adot \period
\end{align}
This may be called a composite $b$-ghost\footnote{We observe that 
 this is very similar to the composite $b$ ghost 
 constructed in the recent ``non-minimal" (topological) formulation 
 of the PS formalism~\cite{BerkTopol} 
and is expected to play an important
 role in deriving that theory from the first principle.}  in the sense that it satisfies 
$\acom{\Qhat}{b_B} =T$ and  $\acom{b_B}{b_B} =0$ just like $b$. 
Then by a similarity transformation $e^X (\star) e^{-X}$ with $X=b_B c$, 
one can remove $T$ and $c$ and obtains 
\begin{align}
e^X \Qhat e^{-X} &= 
\Qtil + {2\over \pplus} \lamtil_\adot \lamtil_\adot b  \comma \\
\Qtil &= \lamtil_a D_a + \lam_\adot D_\adot \comma 
\end{align}
where $\lam_\adot\equiv \lamtil_\adot -(1/2) l_\adot \lamtil_\bdot \lamtil_\bdot$ satisfies the relation  $\lam_\adot 
 \lam_\adot =0$,  recognized as 
 a part of the PS condition. Further, since $c$ 
 is no longer present,
 the last term containing $b$ can be dropped without changing
 the cohomology. Note that even though $T,b,c$ have disappeared we still 
have the relation $\acom{\Qtil}{b_B} =T$. Finally, one can show
 that the non-covariant fermionic fields $S_a$ in $\Qtil$ cohomologically 
 decouple together with 4 of the 8 components of $\lamtil_a$ in such 
 a way that the remaining 11 components of $\lamtil^\al$ precisely form
 a pure spinor $\lam^\al$ satisfying the condition $\lam \ga^m \lam =0$. 
This can again be effected by a similarity transformation~\cite{AK4}  but we will 
 not reproduce the detail here. 
\subsection{Superstring}
The basic idea described above turned out to work for 
 the superstring case as well. However, there were several new complications,
which we list below and briefly describe how they were overcome.
\begin{itemize}
	\item First,  the basic Green-Schwarz action, with 
 a modification of the form (\ref{modification}), is more non-linear
 due to the presence of  the Wess-Zumino  term. This and the existence
of the added spinor make
 the separation between the left-moving and the right-moving sectors 
cumbersome. These complications, however, are only technical and do not
 cause essential problems. 
\item Another difference is that the various quantities  
 now contain $\sigma$-derivatives. 
This leads to a more serious problem: The Dirac brackets among the 
original basic fields are no longer canonical. Fortunately, we were 
 able to overcome this difficulty by constructing
 a modified basis in which the (redefined) fields satisfy canonical 
 bracket relations. 
\item The third new feature in the string case is that in order to 
realize the crucial first class constraint algebra quantum mechanically
 one must make modifications due to multiple contractions
 and the normal-ordering of composite operators. Fortunately again, 
 the needed  modifications were minor and could be found systematically. 
\end{itemize}
In the case of the supermembrane, similar complications are expected 
 to arise. We shall see that some of them can be handled in parallel with 
 the string case but some others present qualitatively new problems. 
\section{Double spinor formalism for supermembrane at the classical level}
\label{sec:DSSM}
Having clarified the basic idea of the double spinor formalism, let us 
now apply it to the supermembrane case. 
\subsection{Fundamental action and its symmetries}
\label{subsec:DSSMaction}
Just as in the superparticle and the superstring cases, the fundamental action 
for the double spinor formalism for the supermembrane is obtained from 
 the conventional BST action~\cite{Bergshoeff:1987cm}
 by simple replacements of fields. 
Setting the membrane tension to unity, it reads 
\begin{align}
\label{eqn:PSSMaction}
S &= \int\mathd^{3}\xi(\calL_{K} + \calL_{WZ}), 
\\
\label{eqn:PSSMactionK}
\calL_{K}
 &= -{1\over2}\sqrt{-g}(g^{IJ}\Pi_{I}^{M}\Pi_{J M} - 1),
\\
\label{eqn:PSSMactionWZ}
\calL_{WZ}
 &= 
 -{1\over2}\epsilon^{IJK}W_{JMN}\bigl(
      \Pi_{J}^{M}\Pi_{K M} + \Pi_{J}^{M}W_{K M} + {1\over3}W_{J}^{M}W_{K M}\bigr)
,    
\end{align}
where
\begin{align}
\Theta \equiv \Tilde{\theta} - \theta\comma \qquad
y^{M} \equiv x^{M} - \mathi\Bar{\theta}\Gamma^{M}\Tilde{\theta}
\comma 
\end{align}
and the basic building blocks are defined as 
\begin{align}
\Pi_{I}^{M} &\equiv \del y^{M} - W_{I}^{M},\quad
\\
W_{I}^{M} &\equiv \mathi\Bar{\Theta}\Gamma^{M}\del_{I}\Theta,\quad
W_{I}^{MN}\equiv \mathi\Bar{\Theta}\Gamma^{MN}\del_{I}\Theta
\,.
\end{align}
Our notations and conventions are as follows\footnote{For more details, 
 see Appendix A.}: 
 $\xi^{I}=(t,\sigma^{i})$ ($i=1,2$) stands for the worldvolume coordinate,
 $x^{M}$ ($M=0,\ldots,10$) is the membrane coordinate, 
 and $\Tilde{\theta}_{A}$ and $\theta_{A}$ ($A=1,\ldots,32$) are 
the two species of Majorana spinors. As before $\theta$ is the newly added
 spinor characteristic of this formalism. The worldvolume metric 
 is denoted by $g_{IJ}$ ($I=0,1,2$). 
As for the $\Gamma$-matrices, we employ 
 the $32$-dimensional Majorana representation
 and denote them by $\Gamma^{M}_{AB}$.
The charge conjugation matrix $C$ equals $\Gamma^{0}$
 and is antisymmetric ($C^{T}=-C$).
Other combinations that frequently appear are
  $C\Gamma^{M}$ and $C\Gamma^{MN}\equiv 
C(\Gamma^{M}\Gamma^{N}-\Gamma^{N}\Gamma^{M})/2$,
  which are both symmetric. The Dirac conjugation
 of a spinor is defined by $\Bar{\theta}_{A}\equiv (\theta C)_{A}$. 

The symmetries possessed by the action above are essentially of the 
 same kind as in the superparticle case. 
In particular, the following three fermionic symmetries
 will be important in the subsequent analyses:
\begin{enumerate}
\item Global supersymmetry:
\begin{align}
\delta\Tilde{\theta}_{A} =0,\quad \delta\theta_{A} = \epsilon_{A},\quad 
 \delta x^{M} = \mathi\Bar{\epsilon}\Gamma^{M}\theta
.
\end{align}
Note that the transformation closes within the $(x^{M},\theta_{A})$ sector.
This will allow us to gauge-fix $\thtil$ {\it without} 
breaking this symmetry. 
\item $\kappa$-symmetry:
\begin{align}
\begin{split}
\delta \Theta_{A} &= (1+\Gamma)\kappa(\xi),\quad
\delta y^{M} = \mathi\Bar{\Theta}\Gamma^{M}(1+\Gamma)\kappa(\xi), \\
\delta(\sqrt{-g}g^{IJ})
&= 2\mathi(\Bar{\kappa}(1+\Gamma)\Gamma_{MN}
 \del_{K}\Theta)g^{KI}\epsilon^{JL_{1}L_{2}}
\Pi_{L_{1}}^{M}\Pi_{L_{2}}^{N} \\
&\quad
 + {2\mathi\over3\sqrt{-g}}
   (\Bar{\kappa}\Gamma^{M}\del^{K}\Theta)\Pi_{K,M}
   \epsilon^{IL_{1}L_{2}}\epsilon^{JL_{3}L_{4}} \\
&\qquad
 \times
   (\Pi_{L_{1}}\cdot \Pi_{L_{3}}\Pi_{L_{2}}\cdot \Pi_{L_{4}}
   + \Pi_{L_{1}}\cdot \Pi_{L_{3}} g_{L_{2}L_{4}}
   + g_{L_{1}L_{3}}g_{L_{2}L_{4}}
   )
 \\
&+(I\leftrightarrow J),
\end{split}
\end{align}
where
\begin{align}
\Gamma& \equiv (1/3!\sqrt{-g})\epsilon^{IJK}
\Pi_{I}^{M}\Pi_{J}^{N}\Pi_{K}^{P}
 \Gamma_{MNP}\,,\quad
 \Gamma^{2}=1\quad \text{(on-shell)}.
\end{align}
This is nothing but the standard 
$\kappa$-symmetry~\cite{Bergshoeff:1987cm,BergHamil}
 written in terms of $(y^{M},\Theta_{A})$.
\item New local fermionic symmetry:
\begin{align}
\label{eqn:localsusy}
\delta\Tilde{\theta}_{A}=\delta\theta_{A}=\chi_{A},\quad
 \delta x^{M} = \mathi(\Bar{\chi}\Gamma^{M}\Theta),\quad
 (\delta\Theta_{A}=\delta y^{M}=0)\,,
\end{align}
where $\chi_{A}(\xi)$ is a local fermionic parameter.
Clearly,  one can use this symmetry to gauge-fix $\theta_{A}$ to zero, 
 upon which our action reduces to the conventional BST action. 
On the other hand,  if we keep this local symmetry till the end
 it is expected to lead to 
a first class algebra of constraints, which was the pivotal element 
for deriving the PS formalism for the superparticle and the 
superstring~\cite{AK4}. 
Below we shall investigate
 if this will be the case for the supermembrane as well. 
\end{enumerate}

\subsection{Analysis of constraints}
\label{subsec:DSSMconstr}
As we shall use the Hamiltonian formulation, 
it is most efficient  to employ the ADM decomposition 
 of the worldvolume metric \cite{ADM}, namely, 
 $\mathd s^{2} = -(N\mathd t)^{2} 
 + \gamma_{ij}(\mathd\sigma^{i}+N^{i}\mathd t)(\mathd \sigma^{j}
 + N^{j}\mathd t)$, 
 where $N,N^i$ and $\gamma_{ij}$ are, respectively, 
the lapse, the shift and the spatial 
metric. We will also use the notation $g\equiv \det g_{ij} = N\sqrt{\gamma}$,
 where $\gamma\equiv \det\gamma_{ij}$. 
In terms of these  ADM variables
 our  action can be written as
\begin{align}
\calL_{K}
&= {1\over2}A\Pi_{0}^{M} \Pi_{0M}
 + B^{M}\Pi_{0M}
 + C,
\\
\calL_{WZ}
&= F_{M}\Pi_{0}^{M} + \Tilde{\Phi}_{A}\Dot{\Theta}_{A},
\end{align}
where\footnote{We define
 $X^{2}\equiv X^{M}X_{M}$ and $X\cdot Y\equiv X^{M}Y_{M}$.}
the quantities 
\begin{align}
A &= {\sqrt{\gamma} \over N},\quad
B^{M} = -AN^{i}\Pi_{i}^{M},\quad
C = {A\over2}N^{2}(1-\gamma^{ij}\Pi_{i}\cdot \Pi_{j}) + {B^{2} \over 2A},
\\
F_{M}
 &= \epsilon^{ij}W_{iMN}\Bigl(\Pi_{j}^{N} + {1\over2}W_{j}^{N}\Bigr),
 \quad
(\epsilon^{12}=1),
\end{align}
are bosonic and 
\begin{align}
\begin{split}
\Tilde{\Phi}_{A}
 &= -{\mathi\over2}\epsilon^{ij}(\Bar{\Theta}\Gamma_{MN})_{A}
  \Bigl(\Pi_{i}^{M}\Pi_{j}^{N} + \Pi_{i}^{M}W_{j}^{N}
   + {1\over3}W_{i}^{M}W_{j}^{N}\Bigr) \\
 &\qquad
   + {\mathi\over2}\epsilon^{ij}W_{iMN}
    (\Bar{\Theta}\Gamma^{M})_{A}
    \Bigl(\Pi_{j}^{N} + {2\over3}W_{j}^{N}\Bigr)
\end{split}
\end{align}
 is fermionic.
It is important to note that $\Pi_{i}^{M}$, $F_{M}$, $\Tilde{\Phi}_{A}$,
 $W_{i}^{M}$ and $W_{iMN}$ 
 are invariant under the local fermionic symmetry (\ref{eqn:localsusy}).

We will denote the canonical conjugates to the basic variables 
 $(N,N^{i},\gamma_{ij},x^{M},\Tilde{\theta}_{A},\theta_{A})$
 by $(P,P_{i},P^{ij},k_{M},\Tilde{k}_{A},k_{A})$. 
They are defined in the standard manner such as 
$k_{A}\equiv (\del / \del \Dot{\theta}_{A})\calL$, where for 
 fermions we use left derivatives. 
The Poisson brackets for the fundamental fields are taken as
\begin{align}
\Poissonss[N,P]&=\delta(\sigma-\sigma'),\quad
\Poissonss[N^{i},P_{j}]= \delta^{i}_{j} \delta(\sigma-\sigma'),\\
\Poissonss[\gamma_{ij},P^{k\ell}]
&= 
 \delta^{k\ell}_{ij} \delta(\sigma-\sigma'),\quad
 (\delta^{k\ell}_{ij} \equiv 
 \delta^{k}_{i}\delta^{\ell}_{j} - \delta^{\ell}_{i}\delta^{k}_{j}),
\\
\Poissonss[x^{M},k^{N}]
&=\eta^{MN}\delta(\sigma-\sigma'),
\\
\Poissonss[\Tilde{\theta}_{A},\Tilde{k}_{B}] &=
\Poissonss[\theta_{A},k_{B}] =
- \delta_{AB} \delta(\sigma-\sigma'),
\\
\text{rest} &= 0.
\end{align}

Since $A$, $B$, $C$, $F$ and $\Tilde{\Phi}_{A}$
 do not contain time derivatives, these 
 canonical conjugates  are readily computed
 and some of them lead to the primary constraints.
First, as the action does not contain the time derivative
 of the worldvolume metric, their conjugates 
 $(P,P_{i},P^{ij})$ vanish.
Similarly, the definitions of the fermionic momenta
 $\Tilde{k}_{A}$ and $k_{A}$ lead to the constraints
\begin{align}
\Tilde{D}_{A}
&\equiv \Tilde{k}_{A}
 - \mathi k_{M}(\Bar{\Tilde{\theta}}\Gamma^{M})_{A}
 +\Tilde{\Phi}_{A}
 \approx0
,
\\
D_{A}
&\equiv k_{A}
 + \mathi k_{M}\bigl((2\Bar{\Tilde{\theta}}-\Bar{\theta})\Gamma^{M}\bigr)_{A}
 - \Tilde{\Phi}_{A}
 \approx0 \comma 
\end{align}
where $\approx0$ means weakly zero. The highly complicated 
 quantity $\Phitil_A$ disappears in the sum 
\begin{align}
\Delta_{A}
 \equiv \Tilde{D}_{A} + D_{A}
 = \Tilde{k}_{A} + k_{A} + \mathi k_{M}(\Bar{\Theta}\Gamma^{M})_{A}
\,,
\label{eqn:genlocalsym}
\end{align}
which can be identified as the generator of the 
local fermionic symmetry.
Below, we shall take $\Tilde{D}_{A}$ and $\Delta_{A}$
 as the independent set of constraints,
 instead of $\Tilde{D}_{A}$ and $D_{A}$.
After some algebra, the 
Poisson brackets among $\Tilde{D}_{A}$ and $\Delta_{A}$ are found as 
\begin{align}
\label{eqn:smTilDTilD}
\Poissonss[\Tilde{D}_{A},\Tilde{D}_{B}]
&= \mathi G_{AB}\delta(\sigma-\sigma'),
\\
\label{eqn:smlocalsusy}
\Poissonss[\Delta_{A},\Delta_{B}]
&= \Poissonss[\Delta_{A},\Tilde{D}_{B}] = 0
,
\end{align}
 where $G_{AB}$ is given by 
\begin{align}
\label{eqn:GAB}
G_{AB} &\equiv 2\calK_{M}(C\Gamma^{M})_{AB}
 + \epsilon^{ij}\Pi_{i}^{M}\Pi_{j}^{N} (C\Gamma_{MN})_{AB}
\,.
\end{align}
The first term on the RHS is the counterpart of 
 the operator $\slash{p}_{\alpha\beta}$ in the superparticle case, 
 while the second term is a new structure characteristic of the 
 supermembrane.

Now, as is usual, we must see if the constraints are consistent
 with the time development.
The canonical Hamiltonian is given,
 up to the constraints described above, by
\begin{align}
H &= \int\mathd^{2}\sigma \calH,\quad
\calH
 = {N\over\sqrt{\gamma}}\calT + N^{i}\calT_{i}
\,,
\end{align}
where
\begin{align}
\begin{split}
\calT &\equiv {1\over2}
 \bigl(\calK^{2} + \gamma(\gamma^{ij}\Pi_{i}\cdot \Pi_{j} - 1)\bigr),\quad
\calT_{i}
 \equiv \calK\cdot \Pi_{i}
 \,,
\\
\calK^{M} &\equiv k^{M} - F^{M}
.
\end{split} \label{eqn:calKM}
\end{align}
$\calT$ and $\calT_i$ are the generators of the worldvolume reparametrization. 
Demanding the consistency of
 the vanishing of 
 $(P,P_{i},P^{ij})$
 with the time development,
 we get the secondary constraints
\begin{align}
T^{(0)} &\equiv \calK^{2} + M \,(\approx 2\calT) \approx 0,\quad
T^{(1)}_{i}\equiv \calT_{i} \approx 0,\quad
T^{(2)}_{ij}\equiv \Pi_{i}\cdot \Pi_{j} - \gamma_{ij} \approx0
,
\end{align}
 where $M_{ij}\equiv \Pi_{i}\cdot \Pi_{j},
M \equiv \det M_{ij}$.
Therefore, the total Hamiltonian at this stage consists 
purely of constraints
\begin{align}
\begin{split}
\label{eqn:HT1}
H_{T} &= \int\mathd^{2}\sigma\calH_{T}, \\
\calH_{T}
&= uP + u_{i}P_{i} + u_{ij}P^{ij} \\
&\qquad + u^{(0)}T^{(0)} + u^{(1)}_{i}T^{(1)}_{i} 
 + u^{(2)}_{ij}T^{(2)}_{ij}
 + \Tilde{v}^{\alpha}\Tilde{D}_{\alpha}
 + v^{\alpha}\Delta_{\alpha}
\,,
\end{split}
\end{align}
where the Lagrange multipliers $u$'s and $v$'s are arbitrary functions
 on the phase space. 
We must check if the constraints are maintained in time
 by computing their Poisson brackets with $H_{T}$. It turned out that 
 our system is already consistent 
 and no new constraints arise: With appropriate 
reorganization, all the constraints weakly commute with $H_{T}$. 
This will become evident in the next subsection, where we consider
 the lightcone decomposition of the constraints.

\subsubsection{Separation of first and second class constraints}
Now we must perform the separation of the first and the second class 
 constraints. This is done by (block) diagonalizing the 
 matrix $C_{IJ}$ given by 
\begin{align}\label{eqn:allc}
{C}_{IJ}&\equiv \Poisson[\phi_{I},\phi_{J}],\quad
\phi_{I}=(P,P_{i},P^{ij},T^{(0)},T^{(1)}_{i},T^{(2)}_{ij},
 \Tilde{D}_{\alpha},\Delta_{\alpha}) \comma 
\end{align}
 on the constraint surface $\phi_{I}=0$. After some analysis, we find 
 that the following set of constraints, equivalent to the original ones, 
do the job:
\begin{align}
&P,\quad P_{i}\,,\quad P^{ij}\,,\quad T^{(2)}_{ij}\,,\quad \Delta_{\alpha}\,,
\\
&\Hat{\Tilde{D}}_{A}
\equiv \Tilde{D}_{A} - 4P^{ij}(C\Gamma^{M}\del_{i}\Tilde{\theta})_{A}\Pi_{jM}\,,
\\
&\Hat{T}^{(0)}
\equiv T^{(0)} -2P^{ij} \del_{i}\Pi_{j} \cdot \calK
+\mathi\epsilon^{ij}(\del_{i}\Bar{\Tilde{\theta}}\Gamma^{M}\Hat{\Tilde{D}})\Pi_{jM}\,,
\\
&\Hat{T}^{(1)}_{i}
\equiv T^{(1)}_{i} - 2\gamma_{jk}\del_{i}P^{jk}
 + 2\del_{i}\Theta_{A}\Tilde{D}_{A}
\,.
\end{align}
The brackets among them all vanish on the constraint surface, except 
 the following two:
\begin{align}
\Poissonss[P^{ij},T^{(2)}_{k\ell}]
 &= \delta^{ij}_{k\ell}\delta(\sigma-\sigma'),
\\
\Poissonss[\Hat{\Tilde{D}}_{A},\Hat{\Tilde{D}}_{B}]
 &\approx \mathi G_{AB}\delta(\sigma-\sigma')
\period
\end{align}
$G_{AB}$, as defined in (\ref{eqn:GAB}), has rank 16.
Therefore, $P^{ij}$, $T^{(2)}_{ij}$
 and a half of $\Hat{D}_{A}$ are of second class
 and all the others are of first class. Actually, 
the pair of second class constraints  $(P^{ij},T^{(2)}_{ij})$
 commute with others even in the strong sense. This means that 
 the Dirac bracket on the constraint surface
 $P^{ij}=T^{(2)}_{ij}=0$ is identical to the original Poisson bracket,
and we may ignore them together with their conjugates 
$(\gamma_{ij},P^{ij})$. 
By choosing the gauge in which $N=1$ and $N^{i}=0$,
 $P$ and $P_{i}$ 
 can be disposed of by the same reason. 

Thus we are left with the remaining constraints
\begin{align}
\Tilde{D}_{A}
&= \Tilde{k}_{A} - \mathi k_{M}(\Bar{\Tilde{\theta}}\Gamma^{M})_{A}
 + \Tilde{\Phi}_{A}\,,
\\
\label{eqn:defT0two}
T^{(0)} 
&= \calK^{2} + M 
 + \mathi \epsilon^{ij}(\del_{i}\Bar{\Tilde{\theta}}\Gamma^{M}\Tilde{D})\Pi_{jM}\,,
\\
\label{eqn:defT1two}
T^{(1)}_{i}
&= \calK\cdot \Pi_{i} + 2\del_{i}\Theta_{A}\Tilde{D}_{A}\,,
\\
\Delta_{\alpha} &= \Tilde{k}_{A} + k_{A} + \mathi k_{M}(\Bar{\Theta}\Gamma^{M})_{A}
\,,
\end{align}
 where for simplicity we removed the hats from 
 the modified constraints
 $(\Hat{\Tilde{D}}_{A},\Hat{T}^{(0)},\Hat{T}^{(1)})$.
Except for a half of $\Tilde{D}_{A}$, all of them are first class:
 $T^{(0)}$ and $T^{(1)}_{i}$ generate the 
 worldvolume reparametrizations,
 $\Delta_{\alpha}$ generates the local fermionic symmetry,
 and the first class part of $\Tilde{D}_{A}$ 
 generates the $\kappa$-symmetry
 (\ref{eqn:smTilDTilD}).

We shall now separate the first and the second class
 part of $\Tilde{D}_{A}$  explicitly by defining
 the generator of $\kappa$-symmetry.
This can be done most efficiently by making the following lightcone 
 decomposition. 
We take a basis of spinors in which
 the lightcone chirality operator 
 (\ie\ the $SO(1,1)$ boost charge) given  by
 $\Hat{\Gamma}=\Gamma^{0}\Gamma^{10}$ is diagonal and decompose spinors
into lightcone chiral and anti-chiral components 
 according to their $\Hat{\Gamma}$ eigenvalues:
 $\phi_{A}=(\phi_{\alpha},\phi_{\Dot{\alpha}})$,
 $\Hat{\Gamma}_{\al\be}\phi_{\be}=\phi_{\alpha}$ and 
$\Hat{\Gamma}_{\aldot\bedot}\phi_{\Dot{\beta}}=-\phi_{\Dot{\alpha}}$.
It will be useful to remember that
 $C\Gamma^{\pm}\equiv C(\Gamma^{0}\pm \Gamma^{10})$, with 
 non-vanishing components  
$C\Gamma^{+}_{\aldot\bedot}=-2\delta_{\Dot{\alpha}\Dot{\beta}}, 
 C\Gamma^{-}_{\al\be}=-2\delta_{\alpha\beta}$,  serve essentially as 
 projectors. As for vectors, the decomposition is defined as 
$v^M = (v^+, v^-, v^m), v^\pm \equiv v^0 \pm v^{10}, m=1,\ldots, 9$. 
(Further details of our conventions can be  found 
in Appendix A.)

In this basis,  $ G_{AB}$
 decomposes as
\begin{align}
G_{\alpha\beta}
&= 2(A+B_{m}\gamma^{m})_{\alpha\beta}\,,\quad
G_{\Dot{\alpha}\Dot{\beta}}
= 2(A'+B'_{m}\gamma^{m})_{\Dot{\alpha}\Dot{\beta}}\,,
 \label{eqn:Galbe} \\
G_{\alpha\Dot{\beta}}
&= (C+D_{m}\gamma^{m}+E_{mn}\gamma^{mn})_{\alpha\Dot{\beta}}\,,\quad 
G_{\Dot{\alpha}\beta}
= (C+D_{m}\gamma^{m}-E_{mn}\gamma^{mn})_{\Dot{\alpha}\beta}\comma 
\end{align}
where
\begin{align}
A &= \calK^{+},\quad B^{m} = -\epsilon^{ij}\Pi_{i}^{+}\Pi_{j}^{m}, 
\\
A' &= \calK^{-},\quad 
 {B^{\prime}}^{m} = \epsilon^{ij}\Pi_{i}^{-}\Pi_{j}^{m},
\\
C &= \epsilon^{ij}\Pi_{i}^{-}\Pi_{j}^{+}\,,\quad
D_{m} = 2\calK_{m}\,,\quad
E_{mn} = \epsilon^{ij}\Pi_{im}\Pi_{jn}\,.
\end{align}
Here and hereafter, we assume that
 $A=\calK^{+}$,
 $B\equiv\sqrt{B^{m}B^{m}}$ 
and  $A^{2}-B^{2}$ are non-vanishing. This is the analogue 
 of the usual assumption 
 in the Green-Schwarz superstring
 that the lightcone momentum $k^{+}$ 
 does not vanish.
Now, to construct the $\kappa$-generator, we will need the inverse 
of the matrix $G_{\al\be}$. Consider the quantity 
 $\Hat{u}_{\alpha\beta} \equiv {1\over B}B^{m}\gamma^{m}_{\alpha\beta}$. 
It is a symmetric matrix satisfying 
 $\Hat{u}^{2}= B_{m}B_{n}\gamma^{m}\gamma^{n}/B^{2}=1$, 
 due to the Clifford algebra.
Therefore, the set of expressions of the type
 $(x+y\Hat{u})_{\alpha\beta}$ appearing in $G_{\al\be}$ 
 form an algebraic field. This immediately allows us to compute the 
the inverse as\footnote{%
Note that $G^{-1}_{\alpha\beta}$ differs from the $\alpha\beta$-component
 of the inverse of the full matrix $G_{AB}$.}
\begin{align}
\label{eqn:Ginv}
G^{-1}_{\alpha\beta}
&= {1\over (A^{2}-B^{2})}(A-B\Hat{u})_{\alpha\beta}\period
\end{align}
Now the $\kappa$-generator $\Tilde{K}_{\Dot{\alpha}}$, 
\ie\ the first class part of $\Tilde{D}_{A}$, 
can be identified as the following linear combination 
of the constraints $\Dtil_\al$ and $\Dtil_\aldot$
\begin{align}
\label{eqn:defKtil}
\Tilde{K}_{\Dot{\alpha}}
&\equiv \Tilde{D}_{\Dot{\alpha}}
 - G_{\Dot{\alpha}\delta}(G^{-1})_{\delta\gamma}\Tilde{D}_{\gamma}
\,.
\end{align}
After  lengthy but straightforward computations, 
its Poisson brackets with $\Dtil_\al$ and with itself 
are found as 
\begin{align}
\label{eqn:TilKTilD}
\Poissonss[\Tilde{K}_{\Dot{\alpha}}, \Tilde{D}_{\beta}]
&= (\text{$\Tilde{D}$-terms}) \approx0,
\\
\label{eqn:TilKTilK}
\Poissonss[\Tilde{K}_{\Dot{\alpha}}, \Tilde{K}_{\beta}]
&= \calT_{\aldot\bedot} \deltassp + (\text{$\Tilde{D}$-terms}) \approx0
\comma 
\end{align}
 where 
\begin{align} 
\label{eqn:memcalTalbe}
 \calT_{\aldot\bedot} &\equiv 
\calT \delta_{\Dot{\alpha}\Dot{\beta}}
 + \calT_{m}\gamma^{m}_{\Dot{\alpha}\Dot{\beta}} \comma \\
\label{eqn:memcalT}
\calT &\equiv {2\over \mathi(A^{2}-B^{2})}
  \bigl(AT^{(0)}
   - 2B_{m}\epsilon^{ij}\Pi^{m}_{i}T^{(1)}_{j}
   + C\epsilon^{ij}\Pi^{+}_{i}T^{(1)}_{j}
   \bigr),
  \\
\label{eqn:memcalTm}
\calT_{m}
&\equiv {2\over \mathi(A^{2}-B^{2})}
 \bigl(B_{m}T^{(0)}
   - 2A\epsilon^{ij}\Pi_{im}T^{(1)}_{j}
   + 2\calK_{m} \epsilon^{ij}\Pi^{+}_{i}T^{(1)}_{j}
 \bigr).
\end{align}
Here ``$\Tilde{D}$-terms'' are 
 those which  vanish upon imposing the fermionic constraint 
 $\Tilde{D}_{A}=0$. 
Since both $\calT$ and $\calT_{m}$ are linear combinations of the 
 original bosonic constraints $T^{(0)}$ and $T^{(1)}_{i}$,
 $\Tilde{K}_{\Dot{\alpha}}$ is indeed of first class. 

This completes the classification of the constraints 
 present in our Hamiltonian system.
The net result may be summarized as follows. The constraints 
 are classified as 
\begin{align}
\text{first class:}&\quad
 P,\quad P_{i}\,,\quad \Delta_{A}\,,\quad T^{(0)},\quad T^{(1)}_{i}\,,\quad
  \Tilde{K}_{\Dot{\alpha}}\,,
\\
\text{second class:}&\quad
 P^{ij}\,,\quad T^{(2)}_{ij}\,,\quad \Tilde{D}_{\alpha}\,,
\end{align}
where $T^\supzero, T^\supone_i$ are the slightly redefined  expressions
 displayed in (\ref{eqn:defT0two}), (\ref{eqn:defT1two}) and 
$\Ktil_\aldot$ is given in (\ref{eqn:defKtil}). 
The total Hamiltonian consists entirely of first class constraints 
and is given by 
\begin{align}
H_{T}
&= \int\mathd^{2}\sigma\calH_{T}\,,\\
\calH_{T}
&= uP+u^{i}P_{i} + u_{A}\Delta_{A} 
 +u^{(0)}T^{(0)} + u^{(1)}_{i}T^{(1)}_{i}
 +\Tilde{u}^{\Dot{\alpha}}\Tilde{K}_{\Dot{\alpha}}
 \,.
\end{align}

\subsection{First class algebra in the SLC gauge}
\label{subsec:DSSMSLC}
\subsubsection{Dirac bracket in the SLC gauge}
Up to this point, 
 we have not gauge-fixed any of the local symmetries of the system. 
Now we fix the  $\kappa$-symmetry generated by 
 $\Tilde{K}_{\Dot{\alpha}}$ by imposing the SLC gauge 
\begin{align}
\Gamma^{+}\Tilde{\theta}=0\quad
\Leftrightarrow
\quad 
\Tilde{\theta}_{\Dot{\alpha}}=0
.
\end{align}
This renders the pair $(\Tilde{\theta}_{\Dot{\alpha}}
,\Tilde{K}_{\Dot{\alpha}})$  second class. 
To define the Dirac bracket on the constraint surface\footnote{%
As we remarked earlier, we may simply ignore
 the constraints $(P,P_{i},P^{ij},T^{(2)}_{ij})$ in the subsequent 
analysis.} specified by 
\begin{align}
\phi_{\Tilde{I}}\equiv 
 (\Tilde{D}_{\alpha},\Tilde{K}_{\Dot{\alpha}},\Tilde{\theta}_{\Dot{\alpha}})=0
,
\end{align}
 one must compute the inverse of the matrix
\begin{align}
C_{\Tilde{I}\Tilde{J}} &\equiv \Poissonss[\phi_{\Tilde{I}},\phi_{\Tilde{J}}]
.
\end{align}
From (\ref{eqn:TilKTilD}), (\ref{eqn:TilKTilK})
 and a simple bracket relation 
$\Poissonss[\Tilde{K}_{\Dot{\alpha}},\Tilde{\theta}_{\Dot{\beta}}]
  = -\delta_{\Dot{\alpha}\Dot{\beta}}\delta(\sigma-\sigma')$,
 $C_{\Tilde{I}\Tilde{J}}$ can be  readily computed. Upon imposing 
 $\phi_{\Tilde{I}}=0$,
it reads
\begin{align}
C_{\Tilde{I}\Tilde{J}} &= \bordermatrix{%
 & \mbox{\scriptsize$\Tilde{D}_{\beta}$}
 & \mbox{\scriptsize$\Tilde{K}_{\Dot{\beta}}$} 
 & \mbox{\scriptsize$\Tilde{\theta}_{\Dot{\beta}}$}
 \cr
\mbox{\scriptsize$\Tilde{D}_{\alpha}$}
 & \mathi G_{\alpha\beta}
 & 0 
 & 0 
 \cr
\mbox{\scriptsize$\Tilde{K}_{\Dot{\alpha}}$}
 & 0 
 & \calT_{\Dot{\alpha}\Dot{\beta}}
 & -\delta_{\Dot{\alpha}\Dot{\beta}} 
 \cr
 \mbox{\scriptsize$\Tilde{\theta}_{\Dot{a}}$}
 & 0 
 & -\delta_{\Dot{\alpha}\Dot{\beta}} 
 & 0 
 \cr}\delta(\sigma-\sigma')
,
\end{align}
and its inverse takes the form
\begin{align}
(C^{-1})_{\Tilde{I}\Tilde{J}} = \bordermatrix{%
 & \mbox{\scriptsize$\Tilde{D}_{\beta}$}
 & \mbox{\scriptsize$\Tilde{K}_{\Dot{\beta}}$} 
 & \mbox{\scriptsize$\Tilde{\theta}_{\Dot{\beta}}$}
 \cr
\mbox{\scriptsize$\Tilde{D}_{{\alpha}}$}
 & -\mathi(G^{-1})_{\alpha\beta}
 & 0 
 & 0 
 \cr
\mbox{\scriptsize$\Tilde{K}_{\Dot{\alpha}}$}
 & 0 
 & 0
 & -\delta_{\Dot{\alpha}\Dot{\beta}} 
 \cr
 \mbox{\scriptsize$\Tilde{\theta}_{\Dot{a}}$}
 & 0 
 & -\delta_{\Dot{\alpha}\Dot{\beta}} 
 & -\calT_{\Dot{\alpha}\Dot{\beta}}
 \cr}\delta(\sigma-\sigma')
\period
\end{align}
Thus the  Dirac bracket on the surface $\phi_{\Tilde{I}}=0$ becomes
\begin{align}
\begin{split}
&\DDiracss[A,B] \\
&\equiv \Poissonss[A,B]
 - \int\mathd^{2}\sigma_{1}
      \Poisson[A(\sigma),\phi_{\Tilde{I}}(\sigma_{1})](C^{-1})_{\Tilde{I}\Tilde{J}}(\sigma_{1})
      \Poisson[\phi_{\Tilde{J}}(\sigma_{1}), B(\sigma')]
      \\
&= \Poissonss[A,B]
 + \mathi\int\mathd^{2}\sigma_{1}
    \Poisson[A(\sigma),\Tilde{D}_{\alpha}(\sigma_{1})](G^{-1})_{\alpha\beta}(\sigma_{1})
    \Poisson[\Tilde{D}_{\beta}(\sigma_{1}), B(\sigma')] \\
&\qquad
 +\int\mathd^{2}\sigma_{1}
    \Poisson[A(\sigma),\Tilde{\theta}_{\Dot{\alpha}}(\sigma_{1})]
      \calT_{\Dot{\alpha}\Dot{\beta}}(\sigma_{1})
    \Poisson[\Tilde{\theta}_{\Dot{\beta}}(\sigma_{1}),B(\sigma')]
\\
&\qquad + \int\mathd^{2}\sigma_{1}
    \Bigl(\Poisson[A(\sigma),\Tilde{\theta}_{\Dot{\alpha}}(\sigma_{1})]
    \Poisson[\Tilde{K}_{\Dot{\alpha}}(\sigma_{1}),B(\sigma')] 
\\
&\qquad\qquad\qquad\qquad\qquad\qquad
    +\Poisson[A(\sigma),\Tilde{K}_{\Dot{\alpha}}(\sigma_{1})]
    \Poisson[\Tilde{\theta}_{\Dot{\alpha}}(\sigma_{1}),B(\sigma')]
    \Bigr)
.
\end{split}
\end{align}

Now, using this  bracket, we compute
 the algebra satisfied by 
 the remaining first class constraints
 $(\Delta_{A},T^{(0)},T^{(1)}_{i})$,
 or equivalently, by $(D_{A},T^{(0)},T^{(1)}_{i})$.
As in the case of the superparticle and the superstring, 
we expect 
 that the information of the $\kappa$-symmetry will 
be reflected in  the brackets involving $D_{\Dot{\alpha}}$. 
More explicitly, the Dirac bracket 
 $\DDirac[D_{\Dot{\alpha}},D_{\Dot{\beta}}]$
would produce $\calT_{\aldot\bedot}$ just as in the Poisson bracket
 $\Poisson[\Tilde{K}_{\Dot{\alpha}},\Tilde{K}_{\Dot{\beta}}]$
given in  (\ref{eqn:TilKTilK}). 
To check this, first note that under the Dirac bracket 
$\DDirac[D_A,D_B]$ is equal to $\DDirac[\Delta_A, \Delta_B]$. 
Further, since  both $\Tilde{D}_{\alpha}$ and $\Tilde{K}_{\Dot{\alpha}}$
 as well as $\Delta_{A}$ itself
 are invariant under the local fermionic symmetry, we have 
\begin{align}
\Poisson[\Delta_{A}, \Tilde{D}_{\alpha}]
=\Poisson[\Delta_{A}, \Tilde{K}_{\Dot{\alpha}}] 
= \Poisson[\Delta_{A}, \Delta_{B}] = 0
\period
\end{align}
Using these relations we easily find
\begin{align}
\label{eqn:memDelDel}
\DDiracss[D_{\Dot{\alpha}}, D_{\Dot{\beta}}]
&=
\DDiracss[\Delta_{\Dot{\alpha}}, \Delta_{\Dot{\beta}}]
=
(\calT \delta_{\Dot{\alpha}\Dot{\beta}} + \calT_{m}\gamma^{m}_{\Dot{\alpha}\Dot{\beta}})\delta(\sigma-\sigma'),
\\
\DDiracss[D_{A},D_{\beta}] 
&= \DDiracss[\Delta_{A},\Delta_{\beta}] = 0
\comma 
\end{align}
which are of the expected form. The relation 
 (\ref{eqn:memDelDel}) is quite analogous to the corresponding  
 formula
 $\DDirac[\Tilde{D}_{\Dot{a}},\Tilde{D}_{\Dot{b}}]
  = (-4\mathi/p^{+})T\delta_{\Dot{a}\Dot{b}}$
 for the superparticle, but there is one crucial difference:
While the constraint $T=p^{2}/2$ in the superparticle case
 simply commutes with itself
 under the Dirac bracket,  this is not the case for the 
reparametrization generators 
 $T^{(0)}$ and $T^{(1)}_{i}$ for the supermembrane. 
As this situation is very similar to the one we encountered 
 in the analysis of the PS superstring, 
 perhaps it is useful to recall briefly
 how this issue was resolved in that case.
 
\subsubsection{A brief revisit to the superstring case}
At the corresponding stage in the analysis of the type II 
superstring in the double spinor formalism, we were left with 
 the first class constraints 
  $(\Delta_{\alpha},\Hat{\Delta}_{\alpha},T^{(0)},T^{(1)})$, 
where 
$\Delta_{\alpha}$ and $\Hat{\Delta}_{\alpha}$ are the generators 
of the local fermionic symmetry for the left/right sectors,
 and $T^{(0)}$ and $T^{(1)}$ are the worldsheet reparametrization generators.
The combinations $T\equiv T^{(0)}+T^{(1)}, \Hat{T}\equiv T^{(1)}-T^{(1)}$
 generate  left/right Virasoro algebras and one obtains a nice 
orthogonal split:
\begin{align}
\begin{split}
L \text{ (left): }&\;(\Delta_{\alpha},T),\qquad
R \text{ (right): }\; (\Hat{\Delta}_{\alpha},\Hat{T}),
\\
\Rightarrow\quad \DDirac[L,R]&=0.
\end{split}
\end{align}
Below, we will exclusively deal with the left sector. 
 
In this sector, we obtained the Dirac bracket relation,
analogous to (\ref{eqn:memDelDel}) above,  of the form 
\begin{align}
\label{eqn:strDelDel}
\DDiracss[D_{\Dot{a}},D_{\Dot{b}}]
 &= 4\mathi \calT \delta_{\Dot{a}\Dot{b}}\delta(\sigma-\sigma'),\quad
\calT \equiv T/\Pi^{+}
\comma 
\end{align}
where $\adot,\bdot$ denote the $SO(8)$ anti-chiral indices and 
 $\Pi^+$ is a component of the superinvariant momentum 
$\Pi^m= k^m + \del_\sig x^m + \cdots$, which is assumed non-vanishing 
 as usual. A remarkable fact was that the  operator $\calT$ on the RHS
 turned out to have 
 a much nicer property
 than $T$.
While $T$ generates the non-trivial Virasoro algebra, 
 $\calT$ has vanishing Dirac brackets
 with $D_{\alpha}$ and with itself, precisely because of the presence of
 the denominator $\Pi^+$. 
Moreover, since $\Pi^{+}\ne0$, obviously $\calT$ and $T$ 
impose the same phase space constraint. 
Therefore, even when the algebra of reparametrization 
 is non-trivial, the crucial algebra formed by the first class constraints
 continues to exhibit a very simple structure. We will now demonstrate 
 that this feature persists for the supermembrane case as well. 
%

\subsubsection{Fundamental constraint algebra for the supermembrane}
Let us now return to the supermembrane theory. 
From the experience with the superstring case just reviewed,  we expect
 that, despite their apparent complexities, 
  $\calT$ and $\calT_m$ would commute among themselves and with $D_A$
 under the Dirac bracket. However, the demonstration by direct computations
 requires a considerable amount of work and is unwieldy. Fortunately, 
there is a much more efficient way, making use of the symmetry structure 
 of the theory. 

The crucial observation 
 is that $\Delta_A$, being the generator of the extra local fermionic 
 symmetry, Poisson-commutes with all the quantities, 
 such as $\calK^M, \Pi^M_i, F_M, \Phitil_A, W^M_i, W_{iMN}$, etc., which 
 are invariant under such a symmetry. This then implies 
 that even under the Dirac bracket  $\Delta_A$ commutes with 
such invariants as long as they have vanishing Poisson brackets with 
 $\thtil_\aldot$, \ie as long as they are free of $\ktil_\aldot$.
 In particular, it is easy to check that $\calT$ and $\calT_m$ 
 are such a quantities and hence we deduce 
$\Dcom{\Delta_A}{\calT} =\Dcom{\Delta_A}{\calT_m}=0$.
 But since $\Delta_A =D_A$ 
 in the SLC gauge, this is equivalent to
\begin{align}
\Dcom{D_A}{\calT}= \Dcom{D_A}{\calT_m}=0 
\period \label{DAcalT}
\end{align}
As for the bracket between $\calT_{\aldot\bedot}$'s, we can make use 
 of the representation
\begin{align}
\calT_{\aldot\bedot}(\sig) 
& =\int d^2\rho \Dcom{D_\aldot(\sig)}{D_\bedot(\rho)} \period
\end{align}
Then, 
\begin{align}
\Dcom{\calT_{\aldot\bedot}(\sig)}{\calT_{\gadot\deltadot}(\sigp)}
&= \int d^2\rho \Dcom{\Dcom{D_\aldot(\sig)}{D_\bedot(\rho)}}{\calT_{\gadot\deltadot}(\sigp)} \comma 
\end{align}
and this vanishes by the use of the graded Jacobi identity and 
(\ref{DAcalT}). Moreover, it is clear that 
 $\calT$ and $\calT_m$ can be independently separated from
 $\calT_{\aldot\bedot}$, as the unit matrix 
 and $\ga^m$ are orthogonal under the trace-norm. 
Hence, we get
\begin{align}
\Dcom{\calT}{\calT} &= \Dcom{\calT}{\calT_m} =\Dcom{\calT_m}{\calT_n} 
 =0 \period
\end{align}

We have thus found a pleasing result: {\it The Dirac bracket
algebra of the first class 
 constraints, $\{D_{A}=(D_{\alpha},D_{\Dot{\alpha}}),\; \calT,\;\calT_{m}\}$, that governs the entire classical dynamics 
 of the supermembrane in the double spinor formalism 
is of  simple structure given by }
\begin{align}
\DDirac[D_{\Dot{\alpha}}, D_{\Dot{\beta}}]
&= \calT\delta_{\Dot{\alpha}\Dot{\beta}}
 + \calT_{m}\gamma^{m}_{\Dot{\alpha}\Dot{\beta}}\comma\label{DDT} \\
\mbox{all others} &=0 \period
\end{align}
As in the case of the superstring, this should serve as the platform 
 upon which to develop the pure spinor type formalism.

We must however note that there is one conspicuous difference from 
 the string case: The term $\calT_m \ga^m_{\aldot\bedot}$
 on the RHS of (\ref{DDT}), 
 which comes from the structure $\ep^{ij}\Pi^M_i\Pi^N_j (C\Ga_{MN})_{AB}$
 in $G_{AB}$ (see (\ref{eqn:GAB})), is new for the supermembrane. 
Consequently, the number of bosonic constraints $\calT=\calT_m=0$
 appears to be more than 
that of the original constraints $T^\supone=T^\suptwo_i=0$. 
In Appendix C, we show that nevertheless the phase space defined by 
 these two sets are equivalent under generic conditions. 
This in turn implies that there must be 7 linear relations 
 among the 10 constraints $\calT_\mbar \equiv (\calT_0=\calT, \calT_m)$, 
 namely 
\begin{align}
Z^\mbar_\pbar \calT_\mbar &= 0 \comma \qquad \pbar =1,2,\ldots 7 \comma 
\label{redrel}
\end{align}
where $Z^\mbar_\pbar$ are field-dependent coefficients. Thus, by definition
the above algebra is of reducible type. To learn the order of the 
 reducibility, one must find $Z^\mbar_\pbar$  and study 
 its properties. After some analysis we find 
that (\ref{redrel}) splits into 
\begin{align}
Z_\pbar^0=0 \comma \qquad Z_\pbar^m \calT_m =0 \comma \label{redrel2}
\end{align}
where $Z_\pbar^m$, given explicitly in Appendix D, are linearly 
 independent as seven 9-vectors $(Z_\pbar)^m$. This shows that the reducibility is of first order. 

Due to this reducibility of the algebra, the construction of the associated
 BRST charge will be more involved than for the string case. 
Fortunately, however, there already exists a general theory~\cite{HenTb} 
 to handle such a situation. We will now describe how it can be applied to
 our case.

%

%

\subsection{Construction of the BRST operator}
\label{subsec:DSSMBRST}
As usual, one starts by introducing the ghosts $\eta_{a_0} =(\lamtil_A, \ctil, 
\ctil_m)$ and their conjugate antighosts $\wp_{a_0}=(\omtil_A, \btil, \btil_m)$ 
corresponding to the constraints $G_{a_0}=(D_A, \calT, \calT_m)$. 
$(\lamtil_A, \omtil_A)$ are bosonic, while $(\ctil, \btil)$ and $(\ctil_m, 
\btil_m)$ are fermionic and they are assumed to satisfy 
 the canonical Dirac bracket relations such as 
$\Dcom{\lamtil_A(\sig)}{\omtil_B(\sigp)} =\delta_{AB} \deltassp$, 
 $\Dcom{\ctil(\sig)}{\btil(\sigp)} = \deltassp$, etc. 
What will be the crucial book-keeping device is  {\it the antighost number}, 
$\agnum$, 
 which at this stage is assigned to be 
1 for $(\omtil_A, \btil, \btil_m)$ and 0 for 
 all the others, including the ghosts. On the other hand, the usual 
 ghost number $\gnum$ is taken to be 1 for the ghosts and $-1$ for 
 the antighosts. The basic strategy for constructing the 
BRST operator $Q$ carrying $\gnum=1$
 is to decompose it according to the antighost number
 as $Q =Q_0+Q_1+\cdots$ and determine $Q_n$ order by order by requiring 
that (i) $Q$ is  nilpotent under the Dirac bracket 
and (ii) its cohomology correctly 
 realizes the gauge invariant functions defined on the constrained surface. 
Referring the reader to \cite{HenTb} for the full details and justifications, 
 below we will explicitly describe the procedure for our system. 

At $\agnum =0$, we start with 
\begin{align}
Q_0 &=  \eta_{a_0} G_{a_0}=\lamtil_A D_A + \ctil \calT + \ctil_m \calT_m \period
\end{align}
This is not nilpotent since
\begin{align}
\Dcom{Q_0}{Q_0}  = 
\lamtil_\aldot \lamtil_\aldot \calT + \lamtil_\aldot \ga^m_{\aldot\bedot}
\lamtil_\bedot 
 \calT_{\aldot\bedot}
\equiv 2D_0 \period
\end{align}
To cure this, one adds $Q_1$ carrying $\agnum=1$. Then, 
$\Dcom{Q_0+Q_1}{Q_0+Q_1} = 2(D_0 + \Dcom{Q_0}{Q_1}) + \Dcom{Q_1}{Q_1}$. 
Thus, to realize the nilpotency at $\agnum=0$, we must require 
$D_0 + [\Dcom{Q_0}{Q_1}]_0 =0 $, where $[\Dcom{Q_0}{Q_1}]_0$ denotes 
 the $\agnum=0$ part of $\Dcom{Q_0}{Q_1}$. 
Such a structure can only be produced when the antighosts in $Q_1$ 
 get contracted with the ghosts in $Q_0$ and disappear. 
With this in mind, one defines the nilpotent 
operator $\delta$, {\it acting only on 
 the antighosts}, by $\delta \wp_{a_0} \equiv G_{a_0}$, \ie 
\begin{align}
\delta \omtil_A &= D_A \comma \quad \delta \btil = \calT \comma \quad 
 \delta \btil_m = \calT_m \period
\end{align}
 Then, it is easy to 
see that $[\Dcom{Q_0}{Q_1}]_0= \delta Q_1$ holds and 
 the equation to solve becomes $\delta Q_1 = -D_0$. Since $D_0$ contains 
 no antighosts, the integrability condition $\delta D_0=0$ is trivially 
satisfied. An obvious solution is $Q_1 =-\half( \lamtil_\aldot \lamtil_\aldot \btil + \lamtil \ga^m \lamtil \btil_m )$, 
for which  $Q_0+Q_1$ takes the usual form of the BRST operator, valid if
the algebra is irreducible. For the reducible case at hand, there is an 
 additional solution of the form $\ga_\pbar Z^m_\pbar \btil_m$, where 
 $\ga_\pbar$ are newly introduced coefficient fields,
  called the ghosts for ghosts\footnote{They are generally denoted by 
$\eta_{a_1}$}. Indeed, $\delta 
(\ga_\pbar Z^m_\pbar \btil_m) = \ga_\pbar Z^m_\pbar \delta \btil_m
= \ga_\pbar Z^m_\pbar \calT_m$, which vanishes due to the reducibility 
 relation (\ref{redrel2}). Thus the solution for $Q_1$ is 
\begin{align}
Q_1 &= -\half \left( \lamtil_\aldot \lamtil_\aldot \btil + \lamtil \ga^m \lamtil \btil_m \right) + \ga_\pbar Z_\pbar^m \btil_m \period
\end{align}

We now move on to the analysis at $\agnum=1$. The 
 details are more complicated but  the basic logic 
 is entirely similar. As the $\agnum=0$ part has been removed, we have
$\Dcom{Q_0+Q_1}{Q_0+Q_1} = 2D_1 + (\text{higher order})$, where
 $\agnum=1$ part $D_1$ is given by 
\begin{align}
D_1 &= \half \ga^\pbar (\ctil \btil_m 
\Dcom{\calT}{Z^m_\pbar} + \ctil_n \btil_m \Dcom{\calT_n}{Z^m_\pbar} )
\period \label{Done}
\end{align}
In obtaining this result, we have used the fact that $\Dcom{D_A}{Z^m_\pbar}
 =0$, following from the invariance of $Z^m_\pbar$ under the 
 local fermionic symmetry, and the reducibility relation (\ref{redrel2}). 
Although the actual computation of the commutators appearing
 in the above expression  is cumbersome and has not yet been performed, 
we can proceed further by using the general structure of the algebra.
Noting that $\calT$ and $\calT_m$ commute among themselves, we get
\begin{align}
\calT_m \Dcom{\calT}{Z^m_\pbar} &= \Dcom{\calT}{ \calT_m Z^m_\pbar} =0 
\comma \quad 
\calT_m \Dcom{\calT_n}{Z^m_\pbar} = \Dcom{\calT_n}{ \calT_m Z^m_\pbar} =0
\period
\end{align}
Further, as $\calT_m Z^m_\pbar =0$ are the only linear relations 
 among $\calT_m$, the vanishing relations above imply that 
$\Dcom{\calT}{Z^m_\pbar}$ and $\Dcom{\calT_n}{Z^m_\pbar}$ must 
 be linear combinations of $Z^m_\pbar$. So we must have
\begin{align}
\Dcom{\calT}{Z^m_\pbar} &= A_\pbar^\qbar Z_\qbar^m \comma \qquad 
\Dcom{\calT_n}{Z^m_\pbar} = B_{n \pbar}^\qbar Z^m_\qbar \comma
\end{align}
where $A_\pbar^\qbar$ and $B_{n \pbar}^\qbar$ are some field-dependent
coefficients. Substituting them into $D_1$ it becomes 
\begin{align}
D_1 &= \half \ga_\pbar (\ctil 
 A_\pbar^\qbar + \ctil_n  
B_{n \pbar}^\qbar  )\btil_m  Z_\qbar^m \period
\end{align}
Just as before, we now introduce $Q_2$ to kill this contribution. 
Focusing at the $\agnum =1$ piece of 
$\Dcom{\sum_{n=0}^2 Q_n}{\sum_{n=0}^2 Q_n}$, we obtain the equation
$\delta Q_2 = -D_1$. The integrability condition $\delta D_1=0$ can be
 checked using $\delta \btil_m = \calT_m$ and $\calT_m Z^m_\pbar =0$. 
$Q_2$ can then be constructed by introducing the antighost 
 $\be_\pbar$, which is
conjugate to $\ga_\pbar$
and carries $\agnum=2$, with the definition of $\delta$ on it as 
$\delta \be_\qbar = \btil_m Z^m_\qbar$. Explicitly it is given by
\begin{align}
Q_2 &= \half \ga_\pbar (\ctil 
 A_\pbar^\qbar + \ctil_n  B_{n \pbar}^\qbar  ) \be_\qbar \period
\end{align}

The general theory~\cite{HenTb} guarantees that this process can be 
 continued consistently to higher orders and the final nilpotent
BRST operator  $Q=\sum_n Q_n$  is unique up to canonical transformations
 in the extended phase space. To know at which stage the series actually 
 terminates requires further calculations, which are left for future study. 

\section{Towards quantization}
\label{sec:quant}
In this section, to pave the way for future developments, 
we will analyze the nature of the remaining problems 
to be solved for the proper quantization of the theory. 
\subsection{``Free field" basis}
\label{subsec:ffbasis}
To perform the quantization of the above system 
in a useful way, the first question to ask 
 is whether one can find a ``free field" basis as in the case of the 
superstring. As we will see, this turns out to be a non-trivial problem. 
To clarify the nature of the difficulty, we will first study the 
 case of the ordinary BST formulation without the extra spinor $\th_A$ in 
 the SLC gauge, before tackling the case of the double spinor formalism. 

Based on the experience with the PS superstring, the best strategy is to 
 begin with the  construction of a self-conjugate spinor field $S_\al$ 
satisfying the canonical Dirac bracket relation
\begin{align}
\Dcom{S_\al(\sig)}{S_\be(\sigp)} &= i\delta_{\al\be} \delta(\sig-\sigp)
\label{SS}
\end{align}
from the original spinor $\thtil_\al$ obeying 
\begin{align}
\Dcom{\thtil_\al(\sig)}{\thtil_\be(\sigp)} &= iG_{\al\be}^{-1} \deltassp
\comma \label{thtilthtil}
\end{align}
where $G_{\al\be}$ is defined in (\ref{eqn:Galbe}). In the SLC gauge
 with $\th_A$ set to zero, $G_{\al\be}$ is simplified considerably and 
 becomes 
\begin{align}
G_{\al\be} &= 2 (A +B^m \ga_m)_{\al\be} \comma \\
A &\equiv  k^+\comma \qquad B^m \equiv -\ep^{ij} \del_i x^+ \del_j x^m 
\period
\end{align}
Also, the second class constraint $\Dtil_\al$ reduces to
\begin{align}
\Dtil_\al &= \ktil_\al -{i \over 2} G_{\al\be} \thtil_\be \period
\label{DtilSLC}
\end{align}
To construct $S_\al$, we will need the square root of the matrix 
$G$, namely $V_{\al\be}$ satisfying 
\begin{align}
V_{\al\ga} V_{\ga\be} = G_{\al\be} \comma 
\end{align}
and its inverse. Similarly to the calculation of 
 the inverse of $G_{\al\be}$ given in (\ref{eqn:Ginv}), 
their explicit forms are easily found as
\begin{align}
V_{\al\be} &= \xi + \xiinv B_m \ga^m \comma \\
\Vinv_{\al\be} &= {1\over 2\sqrt{A^2-B^2}} (\xi -\xiinv B_m \ga^m) 
\comma \label{vinvone}\\
\xi &\equiv {1\over \sqrt{2}} (\sqrt{A+B} + \sqrt{A-B} ) \comma 
\qquad B \equiv \sqrt{B^m B_m} \period \label{vinvtwo}
\end{align}
Since $G_{\al\be}$ and $V_{\al\be}$ do not depend on $\thtil_\al$ nor
 $x^-$, we have $\Pcom{G_{\al\be}}{V_{\ga\delta}} =\Pcom{\Dtil_\al}{V_{\be\ga}}  =0$. Furthermore $V_{\al\be}$
 is independent of $\thtil_\aldot$ and $\ktil_\aldot$
 and hence it Poisson-commutes with $\Ktil_\aldot$ and $\thtil_\aldot$. 
Together, this implies the anticommutation relations under the 
 Dirac bracket $\Dcom{V_{\al\be}}{V_{\ga\delta}}
 = \Dcom{V_{\al\be}}{\thtil_\ga} =0$. This then allows us to define $S_\al$
 as 
\begin{align}
S_\al &= V_{\al\be} \thtil_\be \period \label{DefSal}
\end{align}
Indeed this satisfies $\Dcom{S_\al(\sig)}{S_\be(\sigp)}
=V_{\al\ga}(\sig) V_{\be\delta}(\sigp) 
\Dcom{\thtil_\ga(\sig)}{\thtil_\delta(\sigp)} 
= V_{\al\ga} V_{\be\delta} i \Ginv_{\ga\delta} \deltassp=
 i \delta_{\al\be} \deltassp$, which is the desired canonical relation. 

Next we examine the Dirac brackets among 
 the basic bosonic variables $(x^M, p^M)$, which we collectively
 denote as $f$. Since in general 
$\Pcom{f}{G_{\al\be}} \ne 0$ and hence $\Pcom{f}{\Dtil_\al} \ne 0$, 
they no longer satisfy the canonical relations under the Dirac bracket. 
 Furthermore it is easy to check that $\Dcom{f}{S_\al} \ne 0$. 

To find  new variables which 
satisfy the canonical form of Dirac brackets, it is useful to
 examine the structure 
 of $\Dcom{f}{S_\al}$ in detail. Explicitly we have
\begin{align}
\Dcom{f}{S_\al} &= \Pcom{f}{S_\al} -\Pcom{f}{\Dtil_\be}
{1\over i}\Ginv_{\be\ga} \Pcom{\Dtil_\ga}{S_\al} \period
\end{align}
Substituting the definition of $S_\al$ (\ref{DefSal}) 
and the form of $\Dtil_\ga$ given in (\ref{DtilSLC}), 
and using the relation $0 = \Pcom{f}{1} = \Pcom{f}{V} \Vinv 
 + V \Pcom{f}{\Vinv}$, this can be rewritten as
\begin{align}
\Dcom{f}{S_\al} &= \half U_{\al\be} S_\be \comma \label{fS}\\
U_{\al\be} &= \left( \Pcom{f}{V} \Vinv + \Pcom{f}{\Vinv} V\right)_{\al\be}
\period
\end{align}
It is easy to check that $U$ is antisymmetric and $\Pcom{U}{S_\al} =0$. 
The equation (\ref{fS}) then implies  that if we modify $f$ into 
\begin{align}
\tilde{f} &\equiv f - {i \over 4} U_{\al\be} S_\al S_\be \comma 
\end{align}
we can achieve the desired relation $\Dcom{\tilde{f}}{S_\al} =0$. 

Using the explicit form of $V$ and $\Vinv$ one can readily evaluate
$U_{\al\be}$ and hence 
$\tilde{f}$. It turned out that $x^M$ remain intact  while the 
 new momenta, to be denoted by $p^M$,  are given by 
\begin{align}
p^+ &= k^+ \comma \\
p^- &= k^- -{i \over 4}\del_i (\Uhat^i_{\al\be} S_\al S_\be) \comma \\
p^m &= k^m -{i \over 4}\del_i (\Uhat^{mi}_{\al\be} S_\al S_\be)\comma 
\end{align}
where 
\begin{align}
\Uhat^i &\equiv  
-{2\over \xi^2 \sqrt{A^2-B^2}} \chi^{mi} \ga^{mn} B^n \comma \\
\Uhat^{mi} &\equiv
 -{1\over  \xi^2 \sqrt{A^2-B^2}} \chi^i \ga^{mn} B^n \comma \\
\chi^{mi} &\equiv \ep^{ij}\del_j x^m\comma \qquad \chi^i 
\equiv \ep^{ij} \del_j x^+ \period
\end{align}
We note in passing that, since $x^M$ and $k^+$ are unchanged,  $V_{\al\be}$ 
remains unaltered and is independent of $S_\al$. Therefore, the 
 relation (\ref{DefSal}) can be immediately inverted 
as $\thtil_\al = \Vinv_{\al\be} S_\be$. 

It is  tedious but straightforward  to check 
the Dirac brackets among $x^M$ and the new momenta $p^M$.
 The result is quite satisfying: 
Their brackets are  completely canonical. 
Summarizing, in the case of the conventional 
 BST formulation in the SLC gauge, 
we have succeeded in constructing the basis of fields 
obeying the canonical Dirac bracket relations 
\begin{align}
\Dcom{x^M(\sig)}{p_N(\sigp)} &= i \delta^M_N \deltassp 
\comma \quad \Dcom{S_\al(\sig)}{S_\be(\sigp)} = i\delta_{\al\be}
\deltassp \comma \\
 \mbox{all others}&=0 \period
\end{align}

We now turn to the case of the double spinor formalism. The analysis 
 becomes much more difficult mainly due to the fact that, even in 
 the SLC gauge, $\thtil_\al$ remains in $G_{\al\be}$ in bilinear products with
 the new spinor $\th_A$ and causes $\Pcom{\Dtil_\al}{G_{\be\ga}}$
 to be non-vanishing.  
This renders  the crucial Dirac bracket relations 
$\Dcom{V_{\al\be}}{V_{\ga\delta}}
 = \Dcom{V_{\al\be}}{\thtil_\ga} =0$ no longer valid  and 
  hence the relation between $\thtil_\al$ and $S_\al$ 
 cannot be  given simply by $\thtil_\al = \Vinv_{\al\be} S_\be$. 

For this reason, we have not, unfortunately,  been able to find the 
expression of $\thtil_\al$ in terms of the ``free field" $S_\al$ in 
 a closed form. However, we shall give below an evidence of the existence of
 such a ``free field" basis by explicitly constructing $\thtil_\al$ 
as a power expansion in  $S_\al$ up to $\calO(S^3)$. As the calculations are
 quite involved, we will only sketch the procedure and present the result. 

Since all the Dirac brackets to appear will be local, \ie proportional to
 $\deltassp$, we will often omit for simplicity 
the arguments of the fields $\sig$ and $\sigp$ as well as
   the $\deltassp$ factor. 
What we wish  to do is to construct $\thtil_\al$ satisfying 
 $\Dcom{\thtil_\al}{\thtil_\be} = i\Ginv_{\al\be}$ in powers of  
 the free field $S_\al$ obeying $\Dcom{S_\al}{S_\be} =i \delta_{\al\be}$. 
Thus we expand $\thtil_\al$ as\footnote{In what follows,  the Latin 
indices $p,q,r$ etc. run over the same range as the Greek indices
such as  $\al, \be$.}
\begin{align}
\thtil_\al &= T^\supone_{\al p}S_p + \half T^\suptwo_{\al [pq]}S_p S_q 
+ {1\over 3} T^\supthree_{\al [pqr]} S_p S_q S_r +\cdots \comma 
\label{expthtil}
\end{align}
where the coefficients $T$'s, to be determined,  are assumed to (anti-)commute
 with each other and with $S_\al$ under the Dirac bracket and 
the indices in the bracket $[\quad ]$ are totally antisymmetrized.  
In order to compare the right- and the left-hand sides of the 
 equation $\Dcom{\thtil_\al}{\thtil_\be} = i\Ginv_{\al\be}$, we 
need to first expand $\Ginv_{\al\be}$ in powers of $\thtil_\al$ and 
then re-express it in powers of $S_\al$ using (\ref{expthtil}) itself. 
In the SLC gauge, $G_{\al\be}$ consists of the part $g_{\al\be}$ independent
 of $\thtil_\al$ and the remaining part $(G_1)_{\al\be}$ linear in $\thtil_\al$
 in the following way:
\begin{align}
G_{\al\be} &= g_{\al\be} + (G_1)_{\al\be} \comma \\
g_{\al\be} &= 2 (\pplus + b_m \ga^m)_{\al\be} \comma \qquad
(G_1)_{\al\be} = 2 D_{\ga, \al\be} \thtil_\ga \period \label{gG}
\end{align}
In the above, 
$b_m =-\ep^{ij} \pi^+_j \pi^m_j$, where $\pi^M_i$ is the $\thtil$-independent
 part of $\Pi^M_i$. As for $\pplus$ it is a redefinition of
 the momentum $k^+$ given by 
\begin{align}
\pplus &= k^+ -f^+-\del_i (\thtil_\ga f^i_\ga)  \comma 
\end{align}
where $f^+$ and $f^i_\ga$   are  quantities independent of $\thtil_\al$
 which 
appear in the expression of $F^+$ in the SLC gauge 
as $F^+ = f^+ + \thtil_\ga f_\ga + \del_i \thtil_\ga f^i_\ga$. 
We omit their  explicit expressions since we will not need them. 
What is significant  is that  $\pplus$ can be shown to commute
 with $\Dtil_\al$ under the Dirac bracket. For this reason, we will 
treat $\pplus$ as a whole and do not count the $\thtil$ in it 
 as one power of $\thtil$. Lastly,  $D_{\ga,\al\be}$ is the quantity which 
 will play a central role in the following. It appears
 in the basic Poisson bracket relation\footnote{It can be derived 
 using the formulas listed in Appendix B and a use of a Fierz identity.}
\begin{align}
\Pcom{\Dtil_\ga(\sig)}{G_{\al\be}(\sigp)} &= 2 D_{\ga, \al\be}\deltassp
\comma 
\label{DtilG} 
\end{align}
and is given by 
\begin{align}
D_{\ga, \al\be} &= \phi_\ga \delta_{\al\be} 
 + \rho_{m} \ga^m_{\al\be} \comma \\
\phi_\ga &= f_\ga -\del_i f^i_\ga = 
2i\ep^{ij} \delta_{\ga\gadot} \del_i  \th_\gadot \pi^+_j \comma 
\qquad 
\rho_{m} = -2i\ep^{ij} (\ga_m)_{\ga\gadot} \del_i \th_\gadot
\pi^+_j \period
\end{align}
Note that it is linear in $\th_\aldot$ and hence vanishes in the 
 ordinary BST formulation. It satisfies the Jacobi identity
\begin{align}
D_{\ga, \al\be} + D_{\al, \be\ga} + D_{\be, \ga\al} =0 \comma 
\label{DJacobi}
\end{align}
due to the fact that the LHS of (\ref{DtilG}) is proportional to 
 $\Pcom{\Dtil_\ga}{\Pcom{\Dtil_\al}{\Dtil_\be}}$. 

Now we can describe the procedure to determine the coefficients 
$T^\supone, T^\suptwo$ and $T^\supthree$ in (\ref{expthtil}). 
Regarding $G_{\al\be}$ as a matrix, $\Ginv$ can be expanded 
 in powers of $\thtil_\al$ as $\Ginv = \ginv -\ginv G_1 \ginv 
 + \ginv G_1 \ginv G_1 \ginv - \cdots$ and further in powers of $S_\al$
 using (\ref{expthtil}) for $\thtil_\al$ in $G_1$. Then, the left- and 
 the right-hand sides of the equation
 $\Dcom{\thtil_\al}{\thtil_\be} = i\Ginv_{\al\be}$ become
\begin{align}
{1\over i} LHS &= T^\supone_{\al p} T^\supone_{\be p}  
 -\left( T^\supone_{\al p} T^\suptwo_{\be [p\sig]} S_\sig 
+ (\al \leftrightarrow
 \be) \right) \nn\\
& \quad  +\left(T^\supone_{\al p} T^\supthree_{\be [p \rho \sig]} 
+T^\supone_{\be p} T^\supthree_{\al [p \rho \sig]}
- T^\suptwo_{\al [p \rho]} T^\suptwo_{\be [p \sig]}\right) S_\rho S_\sig + \cdots \comma \\
{1\over i} RHS &= \biggl[ \ginv- 2 \ginv D^\ga \ginv  
T^\supone_{\ga \sig} S_\sig  \nn\\
& - \left(\ginv D^\ga \ginv T^\suptwo_{\ga[\rho\sig]} 
 + 4  \ginv D^\ga \ginv D^\delta \ginv T^\supone_{\ga \rho} 
 T^\supone_{\delta\sig}\right) S_\rho S_\sig + \cdots\biggr]_{\al\be}
\comma 
\end{align}
where $D^\ga$ denotes $D_{\ga,\al\be}$ regarded as a matrix. Equating them 
 and comparing the coefficients  at each order in  $S_\al$, we obtain 
 a set of equations to solve for $T^{(n)}$'s. 

At the zero-th order, 
we get $T^\supone_{\al p} T^\supone_{\be p} = \ginv_{\al\be}$, which 
 can be solved as 
\begin{align}
T^\supone_{\al\be} &= \vinv_{\al\be} \comma 
\end{align}
where $\vinv_{\al\be}$ is the ``square-root" of $\ginv_{\al\be}$ satisfying
$\vinv_{\al\ga} \vinv_{\ga\be} = \ginv_{\al\be}$ and is given by 
the previous formula (\ref{vinvone}) for $\Vinv_{\al\be}$, 
with the substitution $A=\pplus, B_m =b_m$. 

At the first order, 
the equation to solve becomes
$\vinv_{\al \rho} T^\suptwo_{\be [\rho \sig]} +\vinv_{\be \rho}
 T^\suptwo_{\al [\rho \sig]}
= 2 (\ginv D^\ga \ginv)_{\al\be}
\vinv_{\ga \sig} $. After some analysis 
 this can be solved as 
\begin{align}
T^\suptwo_{\al [\be \ga] } 
&=- {2\over 3} (\vinv_{\al\kappa}
 \Dtil_{\be, \ga\kappa }  -\vinv_{\al\kappa}
 \Dtil_{\ga,  \be\kappa}  ) \comma \label{Ttwo}\\
\Dtil_{\al, \be\ga} 
&\equiv \vinv_{\al p} \vinv_{\be q} \vinv_{\ga r}
 D_{p, qr} \period
\end{align}
It should be remarked that the solvability of the equation is 
 rather non-trivial due to the required symmetry property of the 
coefficient $T^\suptwo_{\al[\be\ga]}$. In fact in obtaining (\ref{Ttwo}) 
the use of the Jacobi identity (\ref{DJacobi}) was crucial. 

At the second order, the equation to solve becomes
 considerably more  complicated. Nevertheless, after a long analysis, 
 it can be solved to determine $T^\supthree_{\al[\be\ga\delta]}$. 
We  omit the details and summarize 
 the combined result up to $\calO(S^3)$:
\begin{align}
\thtil_\al &= \vinv_{\al\be} S_\be 
 -{2\over 3} \vinv_{\al p} \Dtil_{\be , \ga p} S_\be S_\ga \nn\\
& + {2\over 9} \vinv_{\al p} (\Dtil_{p, \ga q} \Dtil_{\be,\delta q}
 -5\Dtil_{\ga, p q} \Dtil_{\delta, \be q}) S_\be S_\ga S_\delta
+\calO(S^4) \period
\end{align}
Since $S_\al$ is fermionic, this series will terminate at $\calO(S^{16})$ at 
 the worst. Although  it is difficult to guess 
 the closed form expression at the present time,
 the result above strongly suggests 
 the existence of the ``free field" basis for the double spinor formalism
 as well. 
\subsection{Requirements for proper quantization}
In the above, we have shown that in the SLC gauge 
a basis of fields in which the equal-time Dirac brackets among
 them take the canonical 
 ``free field" form can be constructed in closed form 
for the usual BST formulation and gave an evidence
 that it also exists in the double spinor formalism. Assuming 
 that it exists, the usual step for the quantization of the system 
is to replace the Dirac brackets by the quantum (anti-)commutators. 
It is important, however, to recognize that this procedure constitutes 
 only a part of the quantization. Below we discuss the requirements
  for a proper quantization  and briefly explore a possible scheme
 to realize them  for the supermembrane case. 

Logically,  a classical theory does not determine a corresponding quantum 
 system. In addition to  giving the equal-time (anti-)commutation
 relations for the basic fields, one must also specify (i) the operator 
 products between all the fundamental variables including the 
 nature of short-distance singularities and (ii) the way to define 
 the composite operators which have finite matrix elements. 
These specifications themselves must satisfy certain requirements. The most 
 important among them is that they retain the local symmetries governing the 
degrees of freedom of the system. Otherwise anomalies may result and 
 the quantum system becomes inconsistent. Another requirement is that 
 the physical observables should be hermitian. Further, one often 
demands that the global symmetries of the classical theory remain intact. 
It is not known, however, whether these requirements uniquely fix 
 the quantum theory. There may exist more than one set of rules which 
 satisfy all the requirements. In such a case, they define different 
 but consistent quantum extensions of a given classical theory 
and a choice among them  can only be decided by physical experiments. 

In the case where the theory can be treated perturbatively starting 
 from a free Lagrangian, the procedure 
 of constructing a consistent quantum theory from a classical theory 
 has been systematized through many efforts over the years and 
 is now regarded as a textbook matter. It is instructive, however, to 
 recall the reason why this has been possible. 
For a genuine free theory, one not only has the canonical 
 equal time commutation relations between the conjugate 
 fields, such as $\com{\phi(t,\vec{x})}{\Pi_\phi (t,
\vec{x}')} = i \delta (\vec{x} -\vec{x}')$, but also the crucial relations 
between them, like $\Pi_\phi = \del_t \phi$. 
Because of this, the operator products (or the basic correlation functions)
 among the free fields are completely fixed including the singularities 
 at the coincident point. This then allows one to easily define finite 
composite operators by removing such singularities. When one turns on 
 the interactions, infinities arise at the loop level. But since their 
 structures follow from prescribed rules,
 one can find suitable symmetry-preserving
 regularization scheme and perform renormalization in a systematic manner. 

In contrast, the system that we are dealing with does not admit such a 
 perturbative treatment. In fact the information on the 
 dynamics is contained in its entirety in the set of first class constraints 
in the phase space. Even if one finds a basis of fields 
 where conjugate fields enjoy canonical (anti-)commutation relations, 
such as $\Dcom{x^M(t,\sig)}{p_N(t,\sigp)} = \delta^M_N \deltassp$ 
or its quantum replacement, they are not related simply, 
like $p^M = \del_t x^M$, reflecting the fact that they need not be bonafide
 free fields. In this situation,  what is crucial is to define 
the complete operator products of these fields and the prescription to 
 render the composite operators finite {\it in such a way that the the 
 constraint algebra is realized quantum mechanically}. In the case of 
 the superstring studied in \cite{AK4}, it turned out that this was 
 achieved simply by assuming the free-field operator product together 
with the usual radial normal-ordering and adding a few quantum improvement
 terms to the constraints. In this way the free-field postulate of 
 Berkovits was fully justified. In the case of the supermembrane,
 however, it is not obvious that a similar prescription will work. 
We must postulate certain rules and see if they lead to a consistent 
 quantum theory. 

Although this problem has not been solved, we shall present 
a preliminary investigation, which helps clarify the nature of the problem. 
In what follows, we will concentrate on the local consistency
 and ignore possible 
 global issues. Specifically, 
we will consider the case where the worldvolume is of the 
 structure $R\times \Sigma$, where $R$ denotes the non-compact timelike 
 direction and 
$\Sigma$ is a compact  spatial 2-surface admitting a real complete
 orthonormal basis $\{Y_I(\sig)\}$ for the functions on $\Sigma$ with the 
 properties
\begin{align}
\int d^2\sigma Y_I(\sig) Y_J(\sig) = \delta_{IJ} \comma \qquad 
\sum_I Y_I(\sig)Y_I(\sig') = \deltassp\period
\end{align}

Consider first the bosonic field $x^M(t,\sig)$ and its 
conjugate $p^N(t,\sig)$, which are assumed to satisfy the canonical 
 commutation relations
\begin{align}
\com{x^M(t, \sig)}{p^N(t,\sigp)} &= i\eta^{MN} \deltassp \comma \label{comxp}\\
\com{x^M(t, \sig)}{x^N(t,\sigp)} &= \com{p^M(t, \sig)}{p^N(t,\sigp)} =0 
\period \label{comxxpp}
\end{align}
We expand them in the basis above as 
\begin{align}
x^M(t,\sig) &= \sum_I x^M_I(t)Y_I(\sig) \comma \qquad 
p^N(t,\sig) =\sum_J p^N_J(t)Y_J(\sig) \comma 
\end{align}
and further decompose $x^M(t)$ and $p^M(t)$ into positive and negative 
 frequency parts in the following way:
\begin{align}
x^M_I(t) &= {1\over \sqrt{2\Omega}} 
\int_0^\Omega d\omega \left( a^M_I(\omega) e^{i\omega t} 
 + {a^\dagger}^M_I(\omega) e^{-i\omega t} \right) \comma \\
p^M_I(t) &= {1\over \sqrt{2\Omega}} 
\int_0^\Omega d\omega \left( b^M_I(\omega) e^{i\omega t} 
 + {b^\dagger}^M_I(\omega) e^{-i\omega t} \right) \period
\end{align}
Here, $\Omega$ is a cutoff in frequency to make subsequent computations 
 well-defined. It is easy to see that to realize (\ref{comxp}) we must set
\begin{align}
\com{a^M_I(\omega)}{{b^\dagger }^N_J(\omega') } &= 
\com{{a^\dagger}^M_I(\omega) }{b_I^N(\omega') } = i \eta^{MN} \delta_{IJ}
\delta(\omega -\omega') \comma \\
\com{a^M_I(\omega)}{b^N_J(\omega') } &=\com{{a^\dagger}^M_I
(\omega)}{{b^\dagger}^N_J(\omega') } = 0\period
\end{align}
Note that the commutators $\com{a}{a^\dagger} , 
\com{a}{a } , 
 \com{b}{b^\dagger } $ and $\com{b}{b} $ are not yet determined because 
 of the lack of explicit relation between $x^M$ and $p^M$. Now further 
imposing (\ref{comxxpp}) we get 
\begin{align}
\com{a^M_I(\omega)}{{a^\dagger}^N_J(\omega')} &= \eta^{MN} f_a(I,J)  \delta(\omega-\omega') \comma \\
\com{b^M_I(\omega)}{{b^\dagger}^N_J(\omega')} &= \eta^{MN} f_b(I,J)  \delta(\omega-\omega')\comma \\
\com{a}{a}  &=\com{b}{b} =0 \comma 
\end{align}
where the only restrictions on $f_a(I,J)$ and $f_b(I,J)$ are that they
 are hermitian and symmetric in $I,J$. Generically they may even depend on 
 the operators $a, a^\dagger, b, b^\dagger$. With respect to the states 
 built on the Fock vacuum $\ket{0} $ characterized  by
 $a\ket{0} = b\ket{0} =0$, one can define finite normal-ordered product 
 of two operators $:AB:$ in the usual way, namely by placing $a$ and $b$
  to the  right of $a^\dagger$ and $b^\dagger$. Then the  product
 of spatially separated operators  can be written as 
\begin{align}
A(t, \sig) B(t, \sigp) &= \underwick{1}{<* A(t,\sig) >*B(t,\sigp)}
+ :A(t, \sig) B(t, \sigp):  \comma 
\end{align}
which should be regarded as the definition of 
  the ``contraction" $\underwick{1}{<* A(t,\sig) >*B(t,\sigp)}$. 
Applied to $x^M$ and $p^M$ in question, we readily get
\begin{align}
\underwick{1}{<* x^M(t, \sig) >*p^N(t, \sigp)}
 &= {i \over 2} \eta^{MN} \deltassp
= \half \com{x^M(t, \sig)}{p^N(t, \sigp)} \label{wickxp}
\comma \\
\underwick{1}{<* x^M(t,\sig) >*x^N(t,\sigp)} &=
 \half \eta^{MN} \sum_{I,J} f_a(I,J) Y_I(\sig) Y_J(\sig') \comma \\
\underwick{1}{<* p^M(t,\sig) >*p^N(t,\sigp)} &=
\half \eta^{MN} \sum_{I,J} f_b(I,J) Y_I(\sig) Y_J(\sig')  \period
\end{align}
This makes it crystal clear that, while the singularity of the product of 
 mutually conjugate fields is canonical,  the one between the non-conjugates 
 is dictated by the functions $f_a(I,J)$ and $f_b(I,J)$. The main task 
 of the proper quantization is to choose these functions so that 
 the algebra of first class constraints (with possible modifications 
 of their explicit forms) is maintained quantum mechanically. 

Although this is a difficult problem, a progress can be made
 if a solution exists within the 
 assumption that these functions can be chosen to be $c$-numbers. 
In this case, the normal-ordered product for more than two fields 
 can similarly be defined by pushing the annihilation operators 
 to the right. This gives the familiar recursive definitions (for 
 bosonic fields) 
\begin{align}
A : B_1 B_2 \cdots B_n: &= :A B_1 B_2 \cdots B_n: 
 + \sum_{i=1}^n \underwick{1}{<* A >*B_i} : B_1 \cdots \check{B}_i 
\cdots B_n : \comma \label{normordone}
\end{align}
where $\check{B}_i$ signifies the removal of $B_i$. In this form 
the fact that the normal ordering removes all possible singularities 
 is manifest. Furthermore, it can be shown, again  recursively,  that 
 the normal-ordered product of any number of hermitian fields is
  hermitian. Therefore, once the construction of the canonical basis 
 of fields is completed, it should be possible to examine the 
quantum algebra of constraints just as in the case of genuine free fields. 
In fact the above 
 $c$-number assumption appears reasonable from the point of view of 
 bose-fermi symmetry. If we repeat the analysis given above 
 for the  canonical fermionic field $S_\al$ satisfying 
 $\acom{S_\al(t, \sig)}{S_\be(t, \sigp)} = \delta_{\al\be} \deltassp$, 
we easily find that the singularity in the operator product must be 
 canonical, namely  $\underwick{1}{<* {S}_\al(t,\sig) >* {S}_\be(t,\sigp)}
 =\half \delta_{\al\be} \deltassp$. This is due to the self-conjugate nature
 of  $S_\al$. Now if the local singularity structure is not drastically 
 altered in the double spinor formalism  compared to the usual 
BST formalism, $S_\al$ and $x^M$ should be related by supersymmetry. 
Then, it would be rather unnatural if the singularity in $x^M(t,\sig)
 x^N(t, \sigp)$ is operator valued while the one for $S_\al(t,\sig)S_\be(t,
\sigp)$ is a $c$-number. 
In any case, whether the quantum constraint algebra can be realized with 
 the assumption above should be examined carefully in a future work. 
\section{Summary and discussions}
In this work we started an attempt towards pure spinor type covariant 
 quantization of the supermembrane in 11 dimensions as an application 
 of the ``double spinor" formalism that we developed previously 
 for the superstring. Starting from a simple generalization 
 of the conventional BST action with doubled spinor degrees of 
 freedom and a new local fermionic symmetry, we carefully analyzed the 
structure of the constraints in the semi-light-cone gauge. 
Although the amount of computation is far greater than in the string case, 
 in the end we found a very simple algebra of first class constraints 
 which governs the entire dynamics of the theory. This demonstrates 
 that at least at the classical level the double spinor formalism 
 works nicely for the supermembrane as well. In order to quantize the 
 theory in a tractable way, one must find the basis in which 
 the fundamental fields obey canonical Dirac bracket relations.
 For the BST formulation in the SLC gauge we were able to 
 construct such a basis in  closed form, while for the double spinor
 formulation we indicated its existence by constructing it in the power 
 series in the canonical fermionic field. Finally we discussed in some 
 detail what are required for the proper quantization and suggested 
 a direction to pursue. 

Although the fact that the structure of the all-important first class algebra 
remained simple even for the supermembrane is remarkable and encouraging, 
clearly much work is needed for its quantization and the subsequent 
 extraction of the ``covariant core" of the BRST operator. 
Besides further developing the type of analysis presented in Sec.~4, 
it would be interesting and instructive to study in detail how our results 
reduce to the case of type IIA superstring upon appropriate
 dimensional reduction. We hope to report on these and related matters
elsewhere. 
\par\bigskip\noindent
{\large\bf Acknowledgment}\par\smallskip\noindent
Y.K. acknowledges T.~Yoneya for a useful discussion on the quantization
 problem. 
The research of Y.A. is supported in part by the JSPS Research Fellowship
 for Young Scientists, while that of 
 Y.K. is supported in part by the 
 Grant-in-Aid for Scientific Research (B) 
No.~12440060 and (C) No.~15540256 from the Japan 
 Ministry of Education, Culture, Sports,  Science and Technology. 
\appendix
\setcounter{equation}{0}
\renewcommand{\theequation}{A.\arabic{equation}}
\section*{Appendix A: \ \ Notations and conventions
}
\label{sec:notation}

We take the 32 dimensional Majorana representation for
 the $SO(10,1)$ gamma matrices $\Gamma^{M}$ and they satisfy
\begin{align}
\{\Gamma^{M}\,,\Gamma^{N}\} = 2\eta^{MN}
.
\end{align}
Our convention for the metric is mostly plus: $\eta=(-,+,\ldots,+)$.
The lightcone decomposition of a vector $v^{M}$ is taken as
\begin{align}
 v^{M}=(v^{\pm},v^{m}),\quad
 v^{\pm}=v^{0}\pm v^{10},\quad
 m=1,\ldots,9.
\end{align}
We also define $\Gamma^{\pm}=\Gamma^{0}\pm \Gamma^{10}$.
The lightcone chirality operator (\ie\ the $SO(1,1)$ boost charge)
 is defined as $\Hat{\Gamma}\equiv \Gamma^{0}\Gamma^{10}$
 and we take a basis of spinors in which
 $\Hat{\Gamma}$ is diagonal.
A 32 dimensional spinor $\phi_{A}$ is decomposed as
 $\phi_{A}=(\phi_{\alpha},\phi_{\Dot{\alpha}})$ ($\alpha,\Dot{\alpha}=1,\ldots,16$)
 according to the $\Hat{\Gamma}$-eigenvalue
 ($\Hat{\Gamma}_{\alpha\beta}\phi_{\beta}=\phi_{\alpha}$
 and $\Hat{\Gamma}_{\Dot{\alpha}\Dot{\beta}}\phi_{\Dot{\beta}}=-\phi_{\Dot{\beta}}$),
 and explicitly, we have
\begin{align}
\Gamma^{0}=\begin{pmatrix}0&1\\-1&0\end{pmatrix},\quad
\Gamma^{m}=\begin{pmatrix}-\gamma^{m}&0\\0&\gamma^{m}\end{pmatrix},\quad
\Gamma^{10}=\begin{pmatrix}0&1\\1&0\end{pmatrix},\quad
\Hat{\Gamma}=\begin{pmatrix}1&0\\0&-1\end{pmatrix}.
\end{align}
Note that this basis differs from the ten dimensional chirality
 basis in which $\Gamma^{10}$ is diagonal.
The charge conjugation matrix $C$ 
 is characterized by the property $C(\Gamma^{\mu})^{T}C^{-1} = \Gamma^{\mu}$ 
 and it coincides with $\Gamma^{0}$.
Some formulas useful to remember are
\begin{align}
C\Gamma^{+}&=
\begin{pmatrix}0 & 0 \\ 0 & -2\delta_{\Dot{\alpha}\Dot{\beta}}\end{pmatrix},\quad
C\Gamma^{-}=
\begin{pmatrix}-2\delta_{\alpha\beta} & 0 \\ 0 & 0\end{pmatrix},\quad
C\Gamma^{+-}=
\begin{pmatrix}0&2\delta_{\alpha\Dot{\beta}}\\2\delta_{\Dot{\alpha}\beta}&0\end{pmatrix},
\\
C\Gamma^{+m}&=
\begin{pmatrix}
0&0\\0&-2\gamma^{m}_{\Dot{\alpha}\Dot{\beta}}
\end{pmatrix},\quad
C\Gamma^{-m}=
\begin{pmatrix}2\gamma^{m}_{\alpha\beta}&0\\0&0\end{pmatrix},\quad
C\Gamma^{mn}=
\begin{pmatrix}0&\gamma^{mn}_{\alpha\Dot{\beta}}\\\gamma^{mn}_{\Dot{\alpha}\beta}&0\end{pmatrix}.
\end{align}
The indices of the $16\times16$ components have the
 following symmetries
\begin{align}
\gamma^{m}_{\alpha\beta}=\gamma^{m}_{\beta\alpha},\quad
\gamma^{m}_{\Dot{\alpha}\Dot{\beta}}
=\gamma^{m}_{\Dot{\beta}\Dot{\alpha}},\quad
\gamma^{mn}_{\alpha\Dot{\beta}}=-\gamma^{mn}_{\Dot{\beta}\alpha}\,,
\end{align}
 and $\gamma^{m}_{\alpha\beta}$ and $\gamma^{m}_{\Dot{\alpha}\Dot{\beta}}$
 obey the $SO(9)$ Clifford algebra.
 
Finally, we note the basic Fierz identity
 in 11 dimensions, 
 which we utilized in many of the calculations in the main text:
\begin{align}
 0 &= (C\Gamma^M)_{AB}(C\Gamma_{MN})_{CD}
   + (C\Gamma^M)_{AC}(C\Gamma_{MN})_{DB}
   + (C\Gamma^M)_{AD}(C\Gamma_{MN})_{BC} \nonumber\\
 + (C\Gamma_{MN})_{AB}(C\Gamma_M)_{CD}
   + (C\Gamma_{MN})_{AC}(C\Gamma_M)_{DB}
   + (C\Gamma_{MN})_{AD}(C\Gamma_M)_{BC}\,.
\end{align}

\setcounter{equation}{0}
\renewcommand{\theequation}{B.\arabic{equation}}
\section*{Appendix B: \ \ List of useful formulas in the SLC gauge}
In order to keep the length of the paper reasonable, 
we had to omit many of the calculations, especially those performed in 
 the SLC gauge, where $\thtil_\aldot=0$.
 To partially compensate this omission, below we will 
collect some useful formulas for the basic quantities in this gauge. 

First, before imposing the gauge condition, the basic building blocks $\Pi^M_i, W^M_i(\Th)$ and $W^{MN}_i(\Th)$ are decomposed as 
\begin{align}
\Pi^M_i &= \del_i x^M -i\del_i(\th C\Ga^M \thtil) -W^M_i(\Th) 
= \pi^M_i -\wtil^M_i + 2\atil^M_i \comma \\
W^M_i(\Th) &= i\Th C\Ga^M \del_i \Th = w^M_i + \wtil^M_i 
 -(a^M_i +\atil^M_i) \comma \\
W^{MN}_i(\Th) &= i\Th C\Ga^{MN} \del_i \Th = w^{MN}_i + \wtil^{MN}_i 
 -(b^{MN}_i +\btil^{MN}_i) \comma 
\end{align}
where the variables in lowercase letter are given by
\begin{align}
\pi^M_i &= \del_ix^M -w^M_i \comma \\
w^M_i &= W^M_i(\th)\comma \qquad \wtil^M_i = W^M_i(\thtil) \comma \\
a^M_i &= i\th C\Ga^M\del_i \thtil\comma \quad 
\atil^M_i = i\thtil C\Ga^M\del_i \th \comma \\
w^{MN}_i &= W^{MN}_i(\th)\comma \qquad \wtil^{MN}_i = W^{MN}_i(\thtil) \\
b^{MN}_i &= i\th C\Ga^{MN}\del_i \thtil\comma \quad 
\btil^{MN}_i = i\thtil C\Ga^{MN}\del_i \th \period
\end{align}
$a^M_i, \atil^M_i, b^{MN}_i$ and $\btil^{MN}_i$ are linear in $\thtil$ 
and $\th$. 

Now we go to the SLC gauge by setting $\thtil_\aldot=0$. 
In the following, we will use the notation $\Acute{\th} \equiv (\th_\aldot), 
\th \equiv \th_\al, \thtil \equiv (\thtil_\al)$. 
Then the non-vanishing light-cone components of the variables in lowercase
 letter are 
\begin{align}
\wtil^-_i &= -2i \thtil \del_i \thtil   \comma \qquad 
\wtil^{-m}_i= 2i \thtil \ga^m \del_i \thtil \comma \\
a^m_i &= i\Acute{\th} \ga^m \del_i \thtil 
\comma \qquad a^-_i = -2i \th \del_i \thtil \comma \\
\atil^m_i &= i\thtil \ga^m \del_i \Acute{\th}
 \comma \qquad \atil^-_i = -2i\thtil \del_i \th \comma \\
b^{+-}_i &= 2i \Acute{\th} \del_i \thtil \comma \quad 
 b^{-m} = 2i \th \ga^m \del_i \thtil \comma \quad 
b^{mn}_i = -i\Acute{\th} \ga^{mn} \del_i \thtil \comma \\
\btil^{+-}_i &= 2i\thtil \del_i \Acute{\th} \comma \quad 
\btil^{-m}_i = 2i \thtil \ga^m \del_i \th \comma \quad 
\btil^{mn}_i = -i\thtil \ga^{mn} \del_i \Acute{\th} \period
\end{align}
In terms of them, the light-cone components of the basic building blocks 
 become
\begin{align}
\Pi^+_i &= \pi^+_i \comma \quad 
\Pi^-_i = \pi^-_i -\wtil^-_i + 2\atil^-_i \comma \quad 
\Pi^m_i = \pi^m_i + 2\atil^m_i \comma \\
W^+_i &= w^+_i \comma \quad 
W^-_i = w^-_i + \wtil^-_i 
 -(a^-_i +\atil^-_i) \comma \quad 
W^m_i = w^m_i  -(a^m_i +\atil^m_i) \comma \\
W^{+-}_i &= w^{+-}_i -(b^{+-}_i +\btil^{+-}_i) \comma \quad 
W^{+m}_i = w^{+m}_i \comma \\
W^{-m}_i &=  w^{-m}_i + \wtil^{-m}_i 
 -(b^{-m}_i +\btil^{-m}_i) \comma \quad 
W^{mn}_i = w^{mn}_i -(b^{mn}_i +\btil^{mn}_i)\period
\end{align}
The quantities which play important roles in the text are 
$G_{\al\be}$ and $\Dtil_\al$ given by 
\begin{align}
G_{\al\be} &= 2(A+B_m \ga^m)_{\al\be} \comma \\
A &= k^+-F^+\comma \quad 
 B_m = -\ep^{ij} \Pi^+_i \Pi_{mj} \comma \\
\Dtil_\al &= \ktil_\al -ik^+\thtil_\al +\Phitil_\al \period
\end{align}
In the SLC gauge, $B_m$ simplifies to 
$B_m= -\ep^{ij} \pi^+_i (\pi^m_j + 2\atil^m_j)$. $F^+$ becomes 
\begin{align}
F^+ &= F^{(0)+} + F^{(1)+} \comma \\
F^{(0)+} &=  -\half \ep^{ij} w^{+-}_i\left( \pi^+_j + \half w^+_j\right)
+ \ep^{ij} w^{+m}_i \left( \pi^m_j+ \half w^m_j\right) \comma \\
F^{(1)+} &= \half \ep^{ij} (b^{+-}_i + \btil^{+-}_i)
 \left( \pi^+_j + \half w^+_j\right)
+ \half \ep^{ij}  w^{+m}_i( 3\atil^m_j -a^m_j) \comma 
\end{align}
where $F^{(0)+}$ is independent of $\thtil$ and $F^{(1)+}$
 is linear in $\thtil$. $\Phitil_\al$ is still complicated. If we split 
 it as 
\begin{align}
\Phitil_\al &= \Phitil_\al^{(A)} + \Phitil_\al^{(B)} \comma \\
\Phitil_\al^{(A)} &=-{i \over 2} \ep^{ij} (\Thbar\Ga^{MN})_\al 
 \left( \Pi_{Mi}\Pi_{Nj} + \Pi_{Mi}W_{Nj} + {1\over 3}
W_{Mi}W_{Nj} \right) \comma \\
 \Phitil_\al^{(B)}&=   {i \over 2} \ep^{ij} W^{MN}_i (\Thbar\Ga_M)_\al 
\left( \Pi_{Nj} + {2 \over 3} W_{Nj} \right) \comma 
\end{align}
the light-cone decompositions for 
$\Phitil_\al^{(A)}$ and $\Phitil_\al^{(B)}$ take the form 
\begin{align}
\Phitil_\al^{(A)} &= -{i\over 2} \ep^{ij} 
\Biggl\{ -\half (\acth_\bedot \delta_{\bedot \al})
\Bigl( 2\Pi^-_i \pi^+_j  + \Pi^-_i w^+_j -\pi^+_i W^-_j
+{2\over 3} W^-_i w^+_j  \Bigr) \nn\\
& + (\Th \ga^m)_\al 
\Bigl( 2\Pi^m_i \pi^+_j  + \Pi^m_i w^+_j -\pi^+_i W^m_j 
 + {2\over 3}  W^m_i w^+_j  \Bigr) \nn\\
&+ (\acth \ga^{mn})_\al 
\Bigl( \Pi^m_i \Pi^n_j + \Pi^m_i W^n_j + {1\over 3} W^m_i W^n_j \Bigr) 
\Biggr\} \comma \\
\Phitil_\al^{(B)} &= {i \over 2} \ep^{ij}\Biggl\{ -\half
\left(   \Th_\al W^{+-}_i +(\acth\ga_m)_\al  W^{-m}_i
\right) \left( \pi^+_j + {2\over 3} w^+_j\right) \nn\\
& +\left(  \Th_\al w^{+n}_i -(\acth\ga_m)_\al W^{mn}_i \right) 
\left( \Pi^n_j + {2\over 3} W^n_j
\right) \nn\\
&-\half (\acth\ga_m)_\al w^{+m}_i 
\left( \Pi^-_j + {2\over 3} W^-_j\right)
\Biggr\} \period
\end{align}
\setcounter{equation}{0}
\renewcommand{\theequation}{C.\arabic{equation}}
\section*{Appendix C: \ \ Equivalence between $\{\calT, \calT_m\}$
 and $\{\Tzero, \Tone_i\}$}

In this appendix, we will show that under generic conditions the 
 10 constraints $\{\calT, \calT_m\}$ are equivalent to the original 
  bosonic constraints $\{T^{(0)}, T_i^{(1)}\}$. 

$\calT$ and $\calT_m$, given in (\ref{eqn:memcalT}) and (\ref{eqn:memcalTm}), can be conveniently written as 
\begin{align}
\calT &= {K \over 2i (A^2-B^2)} \comma  \qquad 
\calT_m = {K_m \over 2i (A^2-B^2)} \comma \\
K &= a^i \Tone_i + b\Tzero \comma \qquad 
K_m =  c^i_m \Tone_i + d_m \Tzero \label{KKm} \comma 
\end{align}
where 
\begin{align}
A &= \calK^+  \comma \qquad  B_m = -\ep^{ij} \Pi^+_i \Pi_{mj} 
\comma \qquad B^2 = B_m B^m \comma  \\
a^i &= 4\ep^{ij} \Pi^+_j (\ep^{kl} \Pi^-_k \Pi^+_l) 
-8B_m \ep^{ij} \Pi^m_j \comma \qquad 
b = -4A \comma \\
c^i_m &= 8\ep^{ij} (\Pi^+_j \calK_m -\calK^+ \Pi_{mj}) \comma \qquad 
d_m = -4B_m \comma 
\end{align}
and $\calK^M$ was given in (\ref{eqn:calKM}). 
As stated previously, we assume $A^2-B^2\ne 0$ and $A \ne 0$. 
Since $T^\supone=T^\suptwo_i=0 \Rightarrow \calT=\calT_m=0$ is trivial, 
 what should be shown is the converse. 

First from the form of $c^i_m$ and the definition of $B_m$, one finds that 
$\Pi^+_i c^i_m = 8A B_m$ and hence $d_m = -4B_m = -(1/2A)\Pi^+_i c^i_m$. 
Putting this into $K_m$ yields 
\begin{align}
2i (A^2 -B^2) \calT_m &= c^i_m \left( \Tone_i -{\Pi^+_i \over 2A} 
\Tzero \right) \period
\end{align}
Now generically the $2\times 9$ matrix  $c^i_m$
 has rank 2 and  $C^{ij} \equiv c^i_m c^j_m$ is invertible. 
This means that the linear combination 
$2i (A^2 -B^2) {C^{-1}}_{ij} c^j_m \calT_m$ is actually equal to 
$ \Tone_i -(1/2A) \Pi^+_i \Tzero$. Thus $\calT_m =0$ implies the following
relation
\begin{align}
\Tone_i = {\Pi^+_i \over 2A} \Tzero  \period \label{relTonezero}
\end{align}
Next we note the simple relation $\Pi^+_i a^i = 8B^2$, which can be easily 
 checked. We can use this relation and (\ref{relTonezero}) to rewrite $K$ into 
 the form 
\begin{align}
K &= a^i \Tone_i + b\Tzero = -{4\over A} (A^2-B^2) \Tzero \period
\end{align}
Setting this to zero yields $\Tzero=0$ and this in turn gives 
$\Tone_i=0$ from (\ref{relTonezero}). Therefore under generic conditions
 the constraints $\calT=0, \calT_m=0$ 
are equivalent to the original constraints $\Tzero =0, \Tone_i=0$. 
\setcounter{equation}{0}
\renewcommand{\theequation}{D.\arabic{equation}}
\section*{Appendix D: \ \ First order reducibility function $Z_\pbar^\mbar$}
In this appendix, we give the explicit form of the first order reducibility
 function $Z^\mbar_\pbar$ satisfying 
\begin{align}
Z^\mbar_\pbar \calT_\mbar =0 \comma  \label{ZcalT} 
\end{align}
where  $\calT_\mbar = (\calT_0=\calT, \calT_m)$. 
From the explicit form of $\calT_\mbar$ given in the Appendix C,
 it is easy to see that 
(\ref{ZcalT}) above is equivalent to 
\begin{align}
&A Z_\pbar^0 + B_m Z_\pbar^m =0 \comma \label{eqone}\\
& a^i Z_\pbar^0 + c^i_m Z_\pbar^m =0 \period \label{eqtwo}
\end{align}
First contract (\ref{eqtwo}) with $\Pi^+_i$. Using the relation 
$\Pi^+_i a^i = 8B^2$ already utilized in the Appendix C and a similar one 
$\Pi^+_i c^i_m = 8AB_m$,  which can be easily checked, we get
\begin{align}
B^2 Z_\pbar^0 + AB_m Z^m_\pbar =0 \period \label{eqthree}
\end{align}
Combining (\ref{eqone}) and (\ref{eqthree}) we deduce
\begin{align}
0 &= {A^2-B^2 \over AB^2} B_m Z_\pbar^m \period
\end{align}
Since  $A^2-B^2\ne 0$, this implies   $B_m Z_\pbar^m =0$ and 
further from (\ref{eqone}) we get $Z_\pbar^0=0$. 

To solve  $B_m Z_\pbar^m =0$ for $Z_\pbar^m$ explicitly, 
recall  that $B_m$ is given by $B_m = -\ep^{ij} \Pi^+_i \Pi_{mj}$. Thus 
the equation can be written as 
\begin{align}
\ep^{ij} \Pi^+_i Y_{j\pbar} =0 \comma 
\end{align}
where $Y_{j\pbar} \equiv \Pi_{mj} Z_\pbar^m $. Its general solution
is given by 
\begin{align}
Y_{j\pbar} &=  C_\pbar \Pi^+_j \comma \qquad C_\pbar=\mbox{arbitrary}
\period
\end{align}
Now if we employ the 9-vector notation $\vec{Z}_\pbar =(Z^m_\pbar)$, 
the relation $Y_{j\pbar} = \Pi_{mj} Z_\pbar^m $ can be written as 
\begin{align}
\vec{Z}_\pbar \cdot \vec{\Pi}_1 &=  \Pi^+_1 \comma \qquad 
\vec{Z}_\pbar  \cdot \vec{\Pi}_2 = \Pi^+_2\period \label{eqfour}
\end{align}
where, without loss of generality, 
 we have absorbed the factor $C_\pbar$ into $\vec{Z}_\pbar$. 
Since there are precisely 7 vectors orthogonal to the 
plane spanned by $\vec{\Pi}_i$, we will denote them by $\vec{X}_\pbar$. 
Then the general solution to (\ref{eqfour})
 is of the form 
$\vec{Z}_\pbar = \vec{W} + \vec{X}_\pbar$, 
where $\vec{W}$ is the unique solution of (\ref{eqfour}) lying in the 
$\vec{\Pi}_1$-$\vec{\Pi}_2$ plane.
 Making the expansion $\vec{W} = \xi_i \vec{\Pi}_i $ and plugging it into 
(\ref{eqfour}), the equation for the coefficients becomes 
$R_{ij} \xi_j =\Pi^+_i $, where $R_{ij} \equiv \vec{\Pi}_i\cdot 
 \vec{\Pi}_j = \Pi^m_i \Pi^m_j$.  The solution is 
\begin{align}
\xi_i &= R^{-1}_{ij} \Pi^+_j\comma \qquad R^{-1}_{ij} 
= {\ep_{ik} \ep_{jl} R_{kl} \over \det R} \period
\end{align}
Thus the complete solution for $Z^\mbar_\pbar$, up to an overall 
 factor for each $\pbar$,  is given by 
\begin{align}
\vec{Z}_\pbar &= \vec{X}_\pbar + R^{-1}_{ij} \Pi^+_j \vec{\Pi}_i 
\comma \qquad \vec{X}_\pbar \cdot \vec{\Pi}_i =0 \\
Z_\pbar^0 &=0 \period
\end{align}

It is easy to see that the reducibility is of first order, \ie there are 
 no further linear relations among the seven vectors $\vec{Z}_\pbar$. 
Indeed  $0=\lam^\pbar \vec{Z}_\pbar =\lam^\pbar \vec{X}_\pbar 
 + (\sum_\pbar \lam^\pbar) R^{-1}_{ij} \Pi^+_j \vec{\Pi}_i$ implies 
 $\lam^\pbar =0$, since $\vec{X}_\pbar$ are orthogonal to $\vec{\Pi}_i$ and are linearly independent. 

\newpage

\end{document}